\documentclass[12pt, a4paper]{article}
\pdfoutput=1
\usepackage{graphicx}
\usepackage{amssymb}
\usepackage{amsmath}
\usepackage{bm}
\usepackage[table,dvipsnames]{xcolor} 
\usepackage{cite}
\usepackage{slashed}
\usepackage{epstopdf}            
\usepackage{epsfig}
\usepackage{here}
\usepackage{comment}
\usepackage{booktabs} 
\usepackage{colortbl} 
\usepackage{wrapfig}
\usepackage{ascmac}
\usepackage{fancybox}  
\usepackage{soul} 
\allowdisplaybreaks

\renewcommand{\thefootnote}{\fnsymbol{footnote}}
\numberwithin{equation}{section} 
\def\beq#1\eeq{\begin{align}#1\end{align}}

\RequirePackage{xspace}
\usepackage{caption}
\definecolor{BlueViolet}{rgb}{0.2, 0.00, 0.7}
\definecolor{Blue}{rgb}{0.15, 0.00, 0.9}
\definecolor{light_blue}{rgb}{0.15, 0.35, 0.95}
\definecolor{kit_green}{rgb}{0
, 0.58823 
, 0.50980 
}
\usepackage[
colorlinks=true, linkcolor=light_blue,citecolor=light_blue,urlcolor=kit_green]{hyperref} 

\begin{document}
\sloppy 
\begin{titlepage}
\begin{center}
\hfill{KEK--TH--2680}\\
\vskip .3in

{\Large{\bf Heavy quark symmetry 
\\ \vskip .05in 
behind $b \to c$ semileptonic sum rule}}\\
\vskip .3in

\makeatletter\g@addto@macro\bfseries{\boldmath}\makeatother

{ 
Motoi Endo$^{\rm (a,b)}$,
Syuhei Iguro$^{\rm (a,c,d)}$, 
Satoshi Mishima$^{\rm (e)}$, 
Ryoutaro Watanabe$^{\rm (f)}$
}
\vskip .3in
$^{\rm (a)}${\it KEK Theory Center, IPNS, KEK, Tsukuba 305--0801, Japan}\\\vspace{4pt}
$^{\rm (b)}${Graduate Institute for Advanced Studies, SOKENDAI, Tsukuba, Ibaraki 305--0801, Japan} \\\vspace{4pt}
$^{\rm (c)}${\it Institute for Advanced Research (IAR), Nagoya University,\\ Nagoya 464--8601, Japan}\\\vspace{4pt}
$^{\rm (d)}${\it Kobayashi-Maskawa Institute (KMI) for the Origin of Particles and the Universe, Nagoya University, Nagoya 464--8602, Japan}\\\vspace{4pt}
$^{\rm (e)}${\it Department of Liberal Arts, Saitama Medical University, Moroyama, Saitama 350-0495, Japan}\\\vspace{4pt}
$^{\rm (f)}${\it Institute of Particle Physics and Key Laboratory of Quark and Lepton Physics (MOE), Central China Normal University, Wuhan, Hubei 430079, China}\\


\end{center}
\vskip .15in

\begin{abstract}
Lepton flavor universality violations in semileptonic $b \to c$ transitions have garnered attention over a decade.
For $R_{H_c}={\rm{BR}}(H_b\to H_c \tau\bar\nu_\tau)/{\rm{BR}}(H_b\to H_c \ell\bar\nu_\ell)$ with $\ell$ being $e,\, \mu$, a sum rule among $R_{D}$, $R_{D^*}$ and $R_{\Lambda_c}$ was proposed to check consistency in the experimental results independently of new physics models. 
We revisit this relation from the perspective of the heavy quark symmetry.
We derive a sum rule holding exactly in the heavy quark limit and clarify how model-dependent corrections are introduced in a realistic situation.

\end{abstract}
\end{titlepage}

\setcounter{page}{1}
\renewcommand{\thefootnote}{\#\arabic{footnote}}
\setcounter{footnote}{0}

\hrule
\tableofcontents
\vskip .2in
\hrule
\vskip .4in


\section{Introduction}
\label{sec:intro}

Revealing physics behind the semileptonic $b \to c$ transitions has been a long-standing problem. 
There are tensions between the experimental results and the standard model (SM) predictions in lepton flavor universality (LFU) violations, $R_{H_c}={\rm{BR}}(H_b\to H_c \tau\bar\nu_\tau )/{\rm{BR}}(H_b\to H_c \ell\bar\nu_\ell )$~\cite{HeavyFlavorAveragingGroupHFLAV:2024ctg}.
The deviations for $R_D$ and $R_{D^*}$, so-called the $R_{D^{(*)}}$ anomaly, have reached the $3$ -- $4\, \sigma$ significance levels~\cite{Iguro:2024hyk}.
Therefore, it is imperative to clarify whether they are real signatures of new physics (NP). 

Recently, a relation, referred to as $b \to c$ semileptonic sum rule, was found to hold among the decay rates of $\bar B\to D\tau\bar\nu_\tau$ and $\bar B\to D^*\tau\bar\nu_\tau$, and $\Lambda_b\to \Lambda_c\tau\bar\nu_\tau$~\cite{Blanke:2018yud, Blanke:2019qrx, Fedele:2022iib},
\begin{align}
    \frac{R_{\Lambda_c}}{R_{\Lambda_c}^{\text{SM}}} = \alpha \frac{R_{D}}{R_{D}^{\text{SM}}} + \beta\frac{R_{D^*}}{R_{D^*}^{\text{SM}}} + \delta_{\Lambda_c} \,.
    \label{eq:R_sum_rule}
\end{align}
In each ratio, the denominator is the SM prediction, and the numerator includes NP contributions. 
The coefficients $\alpha$ and $\beta$ (satisfying $\alpha+\beta =1$) are independent of NP models, and miscellaneous model-dependent NP contributions are encoded in $\delta_{\Lambda_c}$. 
The parameters $\alpha$, $\beta$, and $\delta_{\Lambda_c}$ were evaluated in Refs.~\cite{Blanke:2018yud, Blanke:2019qrx, Fedele:2022iib} and recently in Ref.~\cite{Duan:2024ayo} analyzing uncertainties due to form factor inputs. 
They pointed out that $\delta_{\Lambda_c}$ is insignificant when one considers NP interpretations of the $R_{D^{(*)}}$ anomaly.
Hence, by substituting experimental values into the numerators, the sum rule is useful to check consistency in the experimental results independently of NP models. 
For example, the rule can be applied to predict $R_{\Lambda_c}$ from the measured values of $R_{D}$ and $R_{D^{*}}$.
If the result is inconsistent with the experimental value of $R_{\Lambda_c}$, this is most likely not explained by NP models, but there might be defects in the experimental results or theoretical frameworks. 

Despite its usefulness, the sum rule was found empirically, and its theoretical background is unknown. 
Nobody has clarified what is the {\it exact} relation, {\it i.e.,} that satisfied without miscellaneous contributions $\delta_{\Lambda_c}$, and what give rise to $\delta_{\Lambda_c}$. 
One may expect the heavy quark symmetry~\cite{Isgur:1989vq, Neubert:1993mb} to play an essential role.

Reference \cite{Blanke:2019qrx} argued that, in the heavy quark limit, the inclusive rate of $b\to c\tau\bar\nu_\tau$ may be saturated by $\bar B\to D\tau\bar\nu_\tau$ and $\bar B\to D^*\tau\bar\nu_\tau$ for mesonic decays, and by $\Lambda_b\to \Lambda_c\tau\bar\nu_\tau$ for baryons.\footnote{
The measured rate of $\Lambda_b\to \Lambda_c\tau\bar\nu_\tau$ is not large enough for the inclusive one in reality~\cite{Bernlochner:2022hyz}
}
However, it is still unclear how the sum rule \eqref{eq:R_sum_rule} is derived from this argument,\footnote{
For example, although $\Gamma(\bar B\to D^*\tau\bar\nu_\tau) \approx 3\,\Gamma(\bar B\to D\tau\bar\nu_\tau)$ is satisfied in the heavy quark limit within the SM, it does not hold generally in NP models.
} and no further investigation has been explored.

This paper is devoted towards understanding physics behind the $b \to c$ semileptonic sum rule. 
We examine $\bar B\to D\tau\bar\nu_\tau$ and $\bar B\to D^*\tau\bar\nu_\tau$, and $\Lambda_b\to \Lambda_c\tau\bar\nu_\tau$ in the heavy quark effective theory (HQET). 
Under the heavy quark symmetry, hadronic form factors are governed by single Isgur-Wise (IW) functions~\cite{Isgur:1989vq, Isgur:1990pm}. 
Also, bottomed/charmed hadron masses are approximated by heavy quark mass parameters, safely neglecting QCD-scale and $1/m_{c,b}$ corrections in the heavy quark limit.
We will show that such characteristic features enable us to derive an {\it exact} relation among the decay rates, which we will refer to as a sum rule in the heavy quark limit. 
We then discuss how the equality of the relation is violated in reality, {\it i.e.,} by violations of the heavy quark symmetry of the form factors and breakings of the heavy quark relation in the hadron mass spectra.
We will also argue that phase space integrals of the differential decay rates affect the sum rule. 
Consequently, the $b \to c$ semileptonic sum rule \eqref{eq:R_sum_rule} will be derived.

The rest of the paper is organized as follows.
In Sec.~\ref{sec:formulae} we introduce the operator basis to describe NP contributions and review the decay rates.
In Sec.~\ref{sec:SRinHQL}, we derive a sum rule in the heavy quark limit.
Corrections to the sum rule are discussed in Sec.~\ref{sec:Correction}. 
We also compare our results with the sum rule \eqref{eq:R_sum_rule}. 
Section~\ref{sec:Summary} is devoted to the summary and discussion.

\section{Formulae}
\label{sec:formulae}

\subsection{Operator basis}
\label{sec:Heff}

Let us assume that NP contributes to the $b\to c \tau\bar\nu_\tau$ transitions. 
The weak effective Lagrangian is introduced as 
\begin{align}
 \label{eq:Hamiltonian}
 {\cal {L}}_{\rm{eff}}= -\frac{4  G_FV_{cb}}{\sqrt2}\biggl[ (1+C_{V_L})O_{V_L}+C_{S_L}O_{S_L}+C_{S_
R}O_{S_R}+C_{T}O_{T}\biggl]\,.
\end{align}
The effective operators are defined as
\begin{align}
 &O_{V_L} = (\overline{c} \gamma^\mu P_Lb)(\overline{\tau} \gamma_\mu P_L \nu_{\tau})\,,& 
 &O_{S_L} = (\overline{c}  P_Lb)(\overline{\tau} P_L \nu_{\tau})\,,\notag \\
 &O_{S_R} = (\overline{c}  P_Rb)(\overline{\tau} P_L \nu_{\tau})\,,& \label{eq:operator} 
 &O_{T} = (\overline{c}  \sigma^{\mu\nu}P_Lb)(\overline{\tau} \sigma_{\mu\nu} P_L \nu_{\tau}) \,,&
\end{align}
where $P_L=(1-\gamma_5)/2$, $P_R=(1+\gamma_5)/2$, and $\sigma^{\mu\nu}=(i/2)[\gamma^\mu,\gamma^\nu]$ with a convention $\sigma^{\mu\nu} \gamma_5 = - (i/2) \epsilon^{\mu\nu\rho\sigma} \sigma_{\rho\sigma}$. 
Contributions from NP in high energy scales are encapsulated into the Wilson coefficients $C_X$. 
The normalization factor $4 G_F V_{cb}/\sqrt{2}$ corresponds to the SM value, which is obtained by $C_{X} = 0$ for $X=V_L$, $S_{L,R}$, and $T$.
The operator $O_{V_R}=(\overline{c} \gamma^\mu P_Rb)(\overline{\tau} \gamma_\mu P_L \nu_{\tau})$ is neglected because it is generally subject to additional suppression in LFU violating interactions.
All light neutrinos are assumed to be left-handed.\footnote{See Refs.~\cite{Iguro:2018qzf, Robinson:2018gza, Babu:2018vrl, Mandal:2020htr, Penalva:2021wye, Datta:2022czw} for studies including right-handed neutrinos.} 

\subsection{Decay rates in heavy quark effective theory}
\label{sec:decay_rate}

The analysis of the $B \to D^{(*)}$ transitions is based on Refs.~\cite{Bernlochner:2017jka, Iguro:2020cpg}.
In the HQET, the hadron matrix elements are parametrized as  
\begin{align}
 \langle D | \bar c \gamma^\mu b | B \rangle 
 & = \sqrt{m_B m_D} \big[ h_+ (v+v')^\mu + h_- (v-v')^\mu \big] \,, 
 \notag \\[0.5em]
 \langle D |\bar c b| B \rangle
 & = \sqrt{m_B m_D} (w+1) h_S \,, 
 \notag \\[0.5em]
 \langle D |\bar c \gamma^\mu \gamma_5 b| B \rangle
 & = \langle D |\bar c \gamma_5 b| B \rangle = 0 \,, 
 \notag \\[0.5em]
 \langle D |\bar c \sigma^{\mu\nu} b| B \rangle
 & = -i \sqrt{m_B m_D}\, h_T \big( v^\mu v^{\prime\nu} - v^{\prime\mu} v^\nu  \big) \,, 
 \notag \\[0.5em]
 \langle D^* | \bar c \gamma^\mu b | B \rangle
 & = i \sqrt{m_B m_{D^*}} h_V \varepsilon^{\mu\nu\rho\sigma} \epsilon^*_\nu v'_\rho v_\sigma \,, 
 \notag \\[0.5em]
 \langle D^* | \bar c \gamma^\mu \gamma_5 b | B \rangle 
 & = \sqrt{m_B m_{D^*}} \big[ h_{A_1} (w+1) \epsilon^{*\mu} - (\epsilon^* \cdot v) \left( h_{A_2} v^\mu + h_{A_3} v^{\prime\mu} \right) \big] \,, 
 \notag \\[0.5em]
  \langle D^* | \bar c \gamma_5 b | B \rangle 
 & = -\sqrt{m_B m_{D^*}} (\epsilon^* \cdot v) h_P \,, 
 \notag \\[0.5em]
 \langle D^* |\bar c b| B \rangle
 & = 0 \,, 
 \notag \\[0.5em]
 \langle D^* | \bar c \sigma^{\mu\nu} b | B \rangle 
 & = -\sqrt{m_B m_{D^*}} \varepsilon^{\mu\nu\rho\sigma} 
 \big[ h_{T_1} \epsilon^{*}_\rho (v+v')_\sigma + h_{T_2} \epsilon^{*}_\rho (v-v')_\sigma 
 \notag \\ 
 & \qquad\qquad\qquad\qquad~
 + h_{T_3} (\epsilon^* \cdot v) (v+v')_\rho (v-v')_\sigma \big] \,,
\end{align} 
with $v^\mu = p_B^\mu /m_B$ and $v^{\prime\mu} = p_{D^{(*)}}^\mu /m_{D^{(*)}}$.
For $q^2=(p_B-p_{D^{(*)}})^2$, $w =v \cdot v' = (m_B^2+m_{D^{(*)}}^2-q^2)/(2m_Bm_{D^{(*)}})$ varies in the range of $1 \leq w \leq w_{D^{(*)},\,max}$ with $w_{D^{(*)},\,max} = (m_B^2+m_{D^{(*)}}^2-m_\tau^2)/(2m_Bm_{D^{(*)}})$.
The form factors $h_X$ are functions of $w$ and expressed in the heavy quark limit by the leading order IW function $\xi(w)$ as~\cite{Isgur:1989vq}
\begin{align}
 \label{eq:DFFinHQL}
 & h_+ = h_V = h_{A_1} = h_{A_3} = h_S = h_P = h_T = h_{T_1} = \xi(w) \,, \\
 & h_- = h_{A_2} = h_{T_2} = h_{T_3} = 0 \,. \notag
\end{align}
The function satisfies $\xi(1)=1$.
Departing from the heavy quark limit, the form factors include corrections.
Defining $\hat h_X(w) \equiv h_X(w)/\xi(w)$, they are generally expanded as
\begin{align}
 \hat h_X = \hat h_{X,0} + \frac{\alpha_s}{\pi} \delta \hat h_{X,\alpha_s} + \frac{\bar \Lambda}{2m_b} \delta \hat h_{X,m_b} + \frac{\bar \Lambda}{2m_c} \delta \hat h_{X,m_c} + \left(\frac{\bar \Lambda}{2m_c}\right)^2 \delta \hat h_{X,m_c^2} \,, 
 \label{eq:deltaFF}
\end{align} 
where $\hat h_{X,0} = 1$ for $X = +,V,A_1,A_3,S,P,T,T_1$ and $0$ for $X = -,A_2,T_2,T_3$, denoting the leading order contributions.
Also, $\bar \Lambda$ is a QCD scale.
The corrections $\delta \hat h_X$ are taken into account at $\mathcal{O}(\alpha_s, \bar \Lambda/m_{b,c}, \bar \Lambda^2/m_{c}^2)$ by following Refs.~\cite{Bernlochner:2017jka, Iguro:2020cpg}.

Similarly to the $B \to D^{(*)}$ transitions, the HQET form factors for the $\Lambda_b \to \Lambda_c$ transitions are given by~\cite{Bernlochner:2018bfn}
\begin{align}
 \langle \Lambda_c| \bar c\gamma_\mu b |\Lambda_b\rangle 
 &= \bar u(p',s') \big[ f_1 \gamma_\mu + f_2 v_\mu + f_3 v'_\mu \big] u(p,s) \,, 
 \notag \\[0.5em]
 \langle \Lambda_c| \bar c\gamma_\mu\gamma_5 b |\Lambda_b\rangle 
 &= \bar u(p',s') \big[ g_1 \gamma_\mu + g_2 v_\mu + g_3 v'_\mu \big] \gamma_5\, u(p,s) \,, 
 \notag \\[0.5em]
 \langle \Lambda_c| \bar c\, b |\Lambda_b\rangle 
 &= h'_S\, \bar u(p',s')\, u(p,s)
 \,, \notag \\[0.5em]
 \langle \Lambda_c| \bar c \gamma_5 b |\Lambda_b\rangle 
 &= h'_P\, \bar u(p',s')\, \gamma_5\, u(p,s) \,, 
 \notag \\[0.5em]
 \langle \Lambda_c| \bar c\, \sigma_{\mu\nu}\, b |\Lambda_b\rangle 
 &= \bar u(p',s') \big[ h_1\, \sigma_{\mu\nu}
 + i\, h_2 (v_\mu \gamma_\nu - v_\nu \gamma_\mu)
 + i\, h_3 (v'_\mu \gamma_\nu - v'_\nu \gamma_\mu)
 \notag \\ 
 & \qquad\qquad~
 + i\, h_4 (v_\mu v'_\nu - v_\nu v'_\mu) \big] u(p,s) \,,
\end{align}
where $u(p,s)$ are spinors with momentum $p$ and spin $s$.
Also, $v = p/m_{\Lambda_b}$, $v' = p'/m_{\Lambda_c}$, $w = v \cdot v' = (m_{\Lambda_b}^2 + m_{\Lambda_c}^2 - q^2)/(2m_{\Lambda_b} m_{\Lambda_c})$, and $w_{\Lambda_c,\,max} = (m_{\Lambda_b}^2+m_{\Lambda_c}^2-m_\tau^2)/(2m_{\Lambda_b}m_{\Lambda_c})$ are introduced.
The form factors $f_i$, $g_i$, and $h^{(\prime)}_i$ are functions of $w$ and expressed in the heavy quark limit as~\cite{Isgur:1990pm}
\begin{align}
 \label{eq:LambdaFFinHQL}
 & f_1 = g_1 = h'_S = h'_P = h_1 = \zeta(w) \,, \\
 & f_2 = f_3 = g_2 = g_3 = h_2 = h_3 = h_4 = 0 \,, \notag 
\end{align}
where $\zeta(w)$ is the IW function for ground state baryons, satisfying $\zeta(1)=1$.
Once departing from the heavy quark limit, the $\Lambda_b \to \Lambda_c$ form factors include corrections similar to Eq.~\eqref{eq:deltaFF}. 
They are taken into account at $\mathcal{O}(\alpha_s, \bar \Lambda/m_{b,c}, \alpha_s\bar \Lambda/m_{b,c}, \bar \Lambda^2/m_{c}^2)$ by following Refs.~\cite{Bernlochner:2018kxh, Bernlochner:2018bfn}. 

In the HQET, the differential decay rates are described by the above form factors. 
For $\bar B \to D \tau\bar\nu_\tau$, it is expressed as~\cite{Sakaki:2013bfa}
\begin{align}
 \frac{d\Gamma (\bar B \to D \tau\bar\nu_\tau)}{dw} &= 
 \frac{G_F^2 |V_{cb}|^2 \eta_{\rm EW}^2 m_B^3}{48\pi^3} 
 \hat q_D^2 r_D^2 \sqrt{w^2-1} 
 \left( 1 - \rho^2 \right)^2 
 \notag \\ 
 & \times\bigg\{
 |1 + C_{V_L}|^2 
 \left[ \left( 1 + \frac{1}{2} \rho^2 \right) (H_{V,0}^{D})^2 + \frac{3}{2} \rho^2 (H_{V,t}^{D})^2 \right] 
 \notag \\ 
 & \quad
 + \frac{3}{2} \, |C_{S_L} + C_{S_R}|^2 (H_S^{D})^2 
 + 8 \, |C_T|^2 ( 1 + 2 \rho^2 ) (H_T^{D})^2
 \notag \\ 
 & \quad
 + 3 \, \Re\big[ ( 1 + C_{V_L} ) (C_{S_L} + C_{S_R} )^{*} \big] \rho H_S^{D} H_{V,t}^{D} 
 \notag \\ 
 & \quad
 - 12 \, \Re\big[ ( 1 + C_{V_L} ) C_T^{*} \big] \rho H_{V,0}^{D} H_T^{D} 
 \bigg\} \,,
 \label{eq:DDR_D}
\end{align}
where $\hat q_{D^{(*)}}^2 = q^2/m_B^2 = 1 - 2r_{D^{(*)}}w + r_{D^{(*)}}^2$, $r_{D^{(*)}} = m_{D^{(*)}}/m_B$, and $\rho = m_\tau/\sqrt{q^2}$ are introduced.
Also, $\eta_{\rm EW}$ is an EW correction. 
The hadronic amplitudes $H_i^D$ are functions of the form factors shown explicitly in Appendix~\ref{sec:helicity}.
Similarly, the rate for $\bar B \to D^* \tau\bar\nu_\tau$ is shown as~\cite{Sakaki:2013bfa}
\begin{align}
 & \frac{d\Gamma (\bar B \to D^* \tau\bar\nu_\tau)}{dw} =
 \frac{G_F^2 |V_{cb}|^2 \eta_{\rm EW}^2 m_B^3}{48\pi^3} 
 \hat q_{D^*}^2 r_{D^*}^2 \sqrt{w^2-1} 
 \left( 1 - \rho^2 \right)^2 
 \notag \\ 
 & \qquad
 \times\bigg\{
 |1 + C_{V_L}|^2 \left[ \left( 1 + \frac{1}{2} \rho^2 \right) \big[ ( H_{V,0}^{D^*} )^2 + ( H_{V,+}^{D^*} )^2 + ( H_{V,-}^{D^*} )^2 \big] + \frac{3}{2} \rho^2 ( H_{V,t}^{D^*} )^2 \right] 
 \notag \\ 
 & \qquad\quad
 + \frac{3}{2} \, |C_{S_L} - C_{S_R}|^2 ( H_{P}^{D^*} )^2 
 + 8 \, |C_T|^2 ( 1+ 2 \rho^2 ) \big[ ( H_{T,0}^{D^*} )^2 + ( H_{T,+}^{D^*} )^2 + ( H_{T,-}^{D^*} )^2 \big] 
 \notag \\ 
 & \qquad\quad
 - 3 \, \Re\big[ (1 + C_{V_L} ) (C_{S_L} - C_{S_R} )^{*} \big] \rho H_{P}^{D^*} H_{V,t}^{D^*} 
 \notag \\ 
 & \qquad\quad
 - 12 \, \Re\big[ (1 + C_{V_L}) C_T^{*} \big] \rho \left( H_{V,0}^{D^*} H_{T,0}^{D^*} + H_{V,+}^{D^*} H_{T,+}^{D^*} - H_{V,-}^{D^*} H_{T,-}^{D^*} \right) 
 \bigg\} \,.
 \label{eq:DDR_Ds}
\end{align}
On the other hand, the differential decay rate for $\Lambda_b\to\Lambda_c\tau\bar\nu_\tau$ is given by~\cite{Datta:2017aue, Bernlochner:2018bfn}
\begin{align}
 \label{eq:DDR_Lambda}
 & \frac{d\Gamma(\Lambda_b\to\Lambda_c\tau\bar\nu_\tau)}{dw} =
 \frac{G_{F}^{2}|V_{cb}|^2 \eta_{\rm EW}^2 m_{\Lambda_b}^3}{96 \pi^{3}} 
 \hat q_\Lambda^2 r_\Lambda^2 \sqrt{w^2-1} 
 \left( 1 - \rho^2 \right)^2 
 \\ 
 & \quad
 \times\bigg\{
 |1 + C_{V_L}|^2 \bigg[ \bigg( 1 + \frac{1}{2} \rho^2 \bigg) \big[ (H_{V,0+}^\Lambda)^2 + (H_{V,0-}^\Lambda)^2 + (H_{V,1+}^\Lambda)^2 + (H_{V,1-}^\Lambda)^2  \big]
 \notag \\ 
 & \qquad\qquad\qquad\qquad\quad
 + \frac{3}{2} \rho^2 \big[ (H_{V,t+}^\Lambda)^2 + (H_{V,t-}^\Lambda)^2 \big] \bigg]
 \notag \\ 
 & \qquad
 + 3 \Big[ |C_{S_L}+C_{S_R}|^2 (H_{S}^\Lambda)^2 + |C_{S_L}-C_{S_R}|^2 (H_{P}^\Lambda)^2 \Big]
 \notag \\ 
 & \qquad
 + 8 \, |C_T|^2 (1 + 2\rho^2) \Big[ (H_{T,0+}^\Lambda)^2 + (H_{T,0-}^\Lambda)^2 + (H_{T,1+}^\Lambda)^2 + (H_{T,1-}^\Lambda)^2 \Big]
 \notag \\ 
 & \qquad
 + 3 \, \Re\big[ (1+C_{V_L}) (C_{S_L}+C_{S_R})^* \big] \rho (H_{V,t+}^\Lambda + H_{V,t-}^\Lambda) H_{S}^\Lambda 
 \notag \\ 
 & \qquad
 + 3 \, \Re\big[ (1+C_{V_L}) (C_{S_L}-C_{S_R})^* \big] \rho (H_{V,t+}^\Lambda - H_{V,t-}^\Lambda) H_{P}^\Lambda
 \notag \\ 
 & \qquad\
 + 12 \, \Re\big[(1+C_{V_L}) C_T^*\big] \rho
 \big( H_{V,0+}^\Lambda H_{T,0+}^\Lambda + H_{V,0-}^\Lambda H_{T,0-}^\Lambda + H_{V,1+}^\Lambda H_{T,1+}^\Lambda + H_{V,1-}^\Lambda H_{T,1-}^\Lambda \big)
 \bigg\} \,,
 \notag
\end{align}
with $\hat q_\Lambda^2 = q^2/m_{\Lambda_b}^2 = 1 - 2r_{\Lambda}w + r_{\Lambda}^2$ and $r_{\Lambda} = m_{\Lambda_c}/m_{\Lambda_b}$.

\section{Sum rule in heavy quark limit}
\label{sec:SRinHQL}

Let us study the decay rates in the heavy quark limit, $m_{c,b} \gg \bar \Lambda$.
Since the heavy quark symmetry is restored in the limit, the form factors are expressed by the leading order IW functions, and their corrections are suppressed.
Also, the hadron masses are expressed in the HQET as~\cite{Falk:1992wt, Falk:1992ws, Bernlochner:2018bfn}
\begin{align}
 m_{H_Q} = m_Q + \bar \Lambda + \frac{\Delta m^2}{2m_Q} + \ldots \,,
 \label{eq:mass_HQET}
\end{align}
where $m_Q$ is a heavy quark mass, and $\Delta m^2$ is a parameter of QCD-scale squared. 
In the heavy quark limit, the hadron masses are approximated by the first term on the right-hand side, satisfying a heavy quark relation,
\begin{align}
 m_B=m_{\Lambda_b},~~~
 m_{D}=m_{D^*}=m_{\Lambda_c} \,.
 \label{eq:mass_relation}
\end{align}
Hence, $r_{D} = r_{D^*} = r_{\Lambda} \equiv r$ and $\hat q_{D}^2 = \hat q_{D^*}^2 = \hat q_{\Lambda}^2 = 1 - 2rw + r^2 \equiv \hat q^2$ hold.
It is then convenient to define
\begin{align}
 \mathcal{N} \equiv \frac{G_F^2 |V_{cb}|^2 \eta_{\rm EW}^2 m_B^5}{48\pi^3} 
 \hat q^2 r^3 \sqrt{w^2-1} 
 \left( 1 - \rho^2 \right)^2 \,.
\end{align}

As a result, the differential decay rate for $\bar B \to D \tau\bar\nu_\tau$ is simplified as
\begin{align}
 \frac{d\Gamma (\bar B \to D \tau\bar\nu_\tau)}{dw} &= 
 \mathcal{N} \,
 \bigg\{
 |1 + C_{V_L}|^2 
 \bigg[ \left( 1 + \frac{1}{2} \rho^2 \right) \frac{(1+r)^2 (w^2-1)}{\hat q^2}
 \notag \\ 
 & \qquad\qquad\qquad\qquad\qquad
 + \frac{3}{2} \rho^2 \frac{(1-r)^2(w+1)^2}{\hat q^2} \bigg] 
 \notag \\ 
 & \quad
 + \frac{3}{2} \, |C_{S_L} + C_{S_R}|^2 (w+1)^2 
 + 8 \, |C_T|^2 (1+2\rho^2) (w^2-1) 
 \notag \\ 
 & \quad
 + 3 \, \Re\big[ ( 1 + C_{V_L} ) (C_{S_L} + C_{S_R} )^{*} \big] \frac{\rho (1-r) (w+1)^2}{\sqrt{\hat q^2}}
 \notag \\ 
 & \quad
 + 12 \, \Re\big[ ( 1 + C_{V_L} ) C_T^{*} \big] \frac{\rho (1+r) (w^2-1)}{\sqrt{\hat q^2}} 
 \bigg\} \, \xi(w)^2 \,.
 \label{eq:DDR_HQET_D}
\end{align}
Similarly, the differential decay rate for $\bar B \to D^* \tau\bar\nu_\tau$ becomes
\begin{align}
 \frac{d\Gamma (\bar B \to D^* \tau\bar\nu_\tau)}{dw} &=
 \mathcal{N} \,
 \bigg\{
 |1 + C_{V_L}|^2 \bigg[ \left( 1 + \frac{1}{2} \rho^2 \right) 
 \bigg[ 4w(w+1) 
 + \frac{(1-r)^2(w+1)^2}{\hat q^2} \bigg] 
 \notag \\ 
 & \qquad\qquad\qquad\qquad\qquad
 + \frac{3}{2} \rho^2 \frac{(1+r)^2(w^2-1)}{\hat q^2} \bigg] 
 \notag \\ 
 & 
 + \frac{3}{2} \, |C_{S_L} - C_{S_R}|^2 (w^2-1) 
 \notag \\ 
 &
 + 8 \, |C_T|^2 ( 1+ 2 \rho^2 ) \left[ (w+1)^2 + \frac{4(w+1)(r^2w+w-2r)}{\hat q^2} \right] 
 \notag \\ 
 & 
 - 3 \, \Re\big[ (1 + C_{V_L} ) (C_{S_L} - C_{S_R} )^{*} \big] \frac{\rho(1+r)(w^2-1)}{\sqrt{\hat q^2}}
 \notag \\ 
 & 
 + 12 \, \Re\big[ (1 + C_{V_L}) C_T^{*} \big] \frac{\rho(w+1)[r-w+5(rw-1)]}{\sqrt{\hat q^2}}
 \bigg\} \, \xi(w)^2 \,.
 \label{eq:DDR_HQET_Ds}
\end{align}
Also, the rate for $\Lambda_b\to\Lambda_c\tau\bar\nu_\tau$ is obtained as
\begin{align}
 \frac{d\Gamma(\Lambda_b\to\Lambda_c\tau\bar\nu_\tau)}{dw} &=
 4 \mathcal{N} \,
 \bigg\{
 |1 + C_{V_L}|^2 \bigg[ \bigg( 1 + \frac{1}{2} \rho^2 \bigg) \bigg( 2w + \frac{r^2w+w-2r}{\hat q^2} \bigg)
 \notag \\ 
 & \qquad\qquad\qquad\qquad\qquad
 + \frac{3}{2} \rho^2 \frac{r^2w+w-2r}{\hat q^2} \bigg]
 \notag \\ 
 & \quad
 + \frac{3}{4} \Big[ |C_{S_L}+C_{S_R}|^2 (w+1) + |C_{S_L}-C_{S_R}|^2 (w-1) \Big]
 \notag \\ 
 & \quad
 + 8 \, |C_T|^2 (1 + 2\rho^2) \bigg( w + 2\frac{r^2w+w-2r}{\hat q^2} \bigg)
 \notag \\ 
 & \quad
 + \frac{3}{2} \, \Re\big[ (1+C_{V_L}) (C_{S_L}+C_{S_R})^* \big] \frac{\rho(1-r)(1+w)}{\sqrt{\hat q^2}} 
 \notag \\ 
 & \quad
 + \frac{3}{2} \, \Re\big[ (1+C_{V_L}) (C_{S_L}-C_{S_R})^* \big] \frac{\rho(1+r)(1-w)}{\sqrt{\hat q^2}} 
 \notag \\ 
 & \quad
 - 36 \, \Re\big[(1+C_{V_L}) C_T^*\big] \frac{\rho(1-rw)}{\sqrt{\hat q^2}}
 \bigg\} \, \zeta(w)^2 \,.
 \label{eq:DDR_HQET_Lambda}
\end{align}
They are abbreviated as $\kappa_D = d\Gamma (\bar B \to D \tau\bar\nu_\tau)/dw$, $\kappa_{D^*} = d\Gamma (\bar B \to D^* \tau\bar\nu_\tau)/dw$, and $\kappa_{\Lambda_c} = d\Gamma (\Lambda_b \to \Lambda_c \tau\bar\nu_\tau)/dw$ in the following.

For any $C_{V_L}$, $C_{S_L}$, $C_{S_R}$, $C_T$ as well as $w$, one can find a relation among the differential decay rates holding {\it exactly} in the heavy quark limit,\footnote{
 From the relation, $\kappa_{\Lambda_c} \approx \kappa_{D} + \kappa_{D^*}$ is satisfied for $w \approx 1$.
}
\begin{align}
 \frac{\kappa_{\Lambda_c}}{\zeta(w)^2} = \frac{2}{1+w} \frac{\kappa_{D} + \kappa_{D^*}}{\xi(w)^2} \,.
\end{align}
Then, we construct a sum rule by deforming this relation as
\begin{align}
 \frac{\kappa_{\Lambda_c}}{\kappa_{\Lambda_c}^{\rm SM}} = a_{\rm HQL} \frac{\kappa_{D}}{\kappa_{D}^{\rm SM}} + b_{\rm HQL} \frac{\kappa_{D^*}}{\kappa_{D^*}^{\rm SM}} \,.~~~
 (\text{heavy quark limit})
 \label{eq:HQ_sum_rule}
\end{align}
Let us refer to Eq.~\eqref{eq:HQ_sum_rule} as a sum rule in heavy quark limit.
The denominators $\kappa_{H_c}^{\rm SM}$ are the SM values, {\it i.e.}, $C_{V_L} = C_{S_L} = C_{S_R} = C_T = 0$.
The coefficients are given by
\begin{align}
 \label{eq:ab_HQL}
 a_{\rm HQL} &= \frac{2}{1+w} \frac{\zeta(w)^2}{\xi(w)^2} \frac{\kappa_{D}^{\rm SM}}{\kappa_{\Lambda_c}^{\rm SM}} = \frac{(2+\rho^2)(1+r)^2(w-1) + 3\rho^2(1-r)^2(w+1)}{4(1+2\rho^2)(r^2w+w-2r)+ 4(2+\rho^2) w \hat q^2} \,, \\
 b_{\rm HQL} &= \frac{2}{1+w} \frac{\zeta(w)^2}{\xi(w)^2} \frac{\kappa_{D^{*}}^{\rm SM}}{\kappa_{\Lambda_c}^{\rm SM}} = \frac{(2+\rho^2) \Big[ (1-r)^2(w+1) + 4w\hat q^2 \Big] + 3\rho^2(1+r)^2(w-1)}{4(1+2\rho^2)(r^2w+w-2r)+ 4(2+\rho^2) w \hat q^2} \,.
 \notag 
\end{align}
They are independent of the IW functions $\xi(w)$ and $\zeta(w)$, and satisfy $a_{\rm HQL}+b_{\rm HQL}=1$ as the sum rule should hold in the SM limit. 
Since $a_{\rm HQL}$ and $b_{\rm HQL}$ are independent of NP contributions $C_{V_L}$, $C_{S_L}$, $C_{S_R}$, and $C_T$, the sum rule holds in any NP model. 

In Eq.~\eqref{eq:R_sum_rule}, the sum rule consists of ratios of the LFU ratio, {\it i.e.,} double ratios of the decay rates. 
In contrast, Eq.~\eqref{eq:HQ_sum_rule} depends on single ratios of the decay rates,  $\kappa_{H_c}/\kappa_{H_c}^{\rm SM}$. 
It is noticed that decay rates of light lepton channels, $\kappa_{H_c}^\ell = d\Gamma (H_b \to H_c \ell\bar\nu_\ell)/dw$ with $\ell$ being $e,\, \mu$, do not affect the sum rule, because we assume that NP contributes only to $b\to c \tau\bar\nu_\tau$, and $\kappa_{H_c}^\ell = \kappa_{H_c}^{\ell,\, {\rm SM}}$ holds.
Thus, the double ratio becomes
\begin{align}
 \frac{\kappa_{H_c}/\kappa_{H_c}^\ell}{\kappa_{H_c}^{\rm SM}/\kappa_{H_c}^{\ell,\, {\rm SM}}} = \frac{\kappa_{H_c}}{\kappa_{H_c}^{\rm SM}} \,.
\end{align}

\section{Corrections to sum rule}
\label{sec:Correction}

\subsection{Hadron mass spectra and form factors}
\label{sec:NonHQL}

Although the sum rule \eqref{eq:HQ_sum_rule} is exact in the heavy quark limit, the realistic bottomed and charmed mass spectra do not satisfy the heavy quark relation \eqref{eq:mass_relation} but involve the corrections in Eq.~\eqref{eq:mass_HQET}. 
Additionally, sub-leading contributions to the form factors \eqref{eq:deltaFF} arise because the heavy quark symmetry is broken. 
Consequently, the equality of the sum rule is violated.

The deviation from the equality is parametrized as 
\begin{align}
 \overline{\delta}_{\Lambda_c} \equiv 
 \frac{\kappa_{\Lambda_c}}{\kappa_{\Lambda_c}^{\rm SM}} - a_{\rm HQL} \frac{\kappa_{D}}{\kappa_{D}^{\rm SM}} - b_{\rm HQL} \frac{\kappa_{D^*}}{\kappa_{D^*}^{\rm SM}} \,,
\label{eq:delta}
\end{align}
where the differential decay rates are no longer evaluated in the heavy quark limit. 
We decompose the decay rates \eqref{eq:DDR_D}, \eqref{eq:DDR_Ds}, and \eqref{eq:DDR_Lambda} by the Wilson coefficients as
\begin{align}
 \kappa_{H_c} \, \left( = \frac{d\Gamma(H_b \to H_c\tau\bar\nu_\tau)}{dw}\right) \equiv \sum_{ij}  \mathcal{C}_i \mathcal{C}_j^*\,\kappa^{ij}_{H_c} \,.
\end{align}
The Wilson coefficient factor $\mathcal{C}_i$ represents
\begin{align}
 \mathcal{C}_i = 
 \begin{cases}
 \, 1+C_{V_L}, & \text{for}~i=V_L \\
 \, C_{S_L}, & \text{for}~i=S_L \\
 \, C_{S_R}, & \text{for}~i=S_R \\
 \, C_T. & \text{for}~i=T
 \end{cases}
\end{align}
Since $\kappa^{S_LS_L}_{H_c} = \kappa^{S_RS_R}_{H_c}$ is satisfied, let us express $(ij) = (S_LS_L)$ and $(S_RS_R)$ to be $(ij) = (SS)$.
Then, Eq.~\eqref{eq:delta} is rewritten as
\begin{align}
 \overline{\delta}_{\Lambda_c} = 
 \sum_{ij}\,
 \mathcal{C}_i \mathcal{C}_j^* 
 \left[ \frac{\kappa^{ij}_{\Lambda_c}}{\kappa^{V_LV_L}_{\Lambda_c}} - a_{\rm HQL} \frac{\kappa^{ij}_{D}}{\kappa_{D}^{V_LV_L}} - b_{\rm HQL} \frac{\kappa^{ij}_{D^*}}{\kappa^{V_LV_L}_{D^*}} \right]  \,,
 \label{eq:delta_mod}
\end{align}
where the denominators express the SM values, since $\kappa_{H_c}^{\rm SM} = \kappa_{H_c}^{V_LV_L}$ holds.

The deviation $\overline{\delta}_{\Lambda_c}$ may be improved by shifting $a_{\rm HQL}$ and $b_{\rm HQL}$ such that some of the $\mathcal{C}_i \mathcal{C}_j^*$ terms are suppressed.\footnote{
Such a prescription was also adopted in Refs.\cite{Blanke:2018yud, Blanke:2019qrx, Fedele:2022iib, Duan:2024ayo} when the sum rule \eqref{eq:R_sum_rule} was given.
}
Instead of Eq.~\eqref{eq:delta_mod}, let us redefine the deviation as
\begin{align}
 \overline{\delta}_{\Lambda_c}^{kl} \equiv \sum_{ij}  \mathcal{C}_i \mathcal{C}_j^* \,\overline{\delta}_{\Lambda_c}^{kl}(ij)\,,
 ~~~
 \overline{\delta}_{\Lambda_c}^{kl}(ij) = 
 \frac{\kappa^{ij}_{\Lambda_c}}{\kappa^{V_LV_L}_{\Lambda_c}} - a_{\kappa}^{kl} \frac{\kappa^{ij}_{D}}{\kappa_{D}^{V_LV_L}} - b_{\kappa}^{kl} \frac{\kappa^{ij}_{D^*}}{\kappa^{V_LV_L}_{D^*}} \,.
 \label{eq:delta_mod_2}
\end{align}
We first require that the sum rule is exact in the SM limit, {\it i.e.,} $\overline{\delta}_{\Lambda_c}^{kl}(V_LV_L) = 0$.
Additionally, $\overline{\delta}_{\Lambda_c}^{kl}(kl) = 0$ is achieved if $a_{\kappa}^{kl}$ and $b_{\kappa}^{kl}$ satisfy
\begin{align}
 a_{\kappa}^{kl} = \frac{\kappa_{\Lambda_c}^{kl}/\kappa_{\Lambda_c}^{V_LV_L}-\kappa_{D^*}^{kl}/\kappa_{D^*}^{V_LV_L}}{\kappa_{D}^{kl}/\kappa_{D}^{V_LV_L}-\kappa_{D^*}^{kl}/\kappa_{D^*}^{V_LV_L}} \,,~~~
 b_{\kappa}^{kl} = \frac{\kappa_{D}^{kl}/\kappa_{D}^{V_LV_L}-\kappa_{\Lambda_c}^{kl}/\kappa_{\Lambda_c}^{V_LV_L}}{\kappa_{D}^{kl}/\kappa_{D}^{V_LV_L}-\kappa_{D^*}^{kl}/\kappa_{D^*}^{V_LV_L}} \,.
 \label{eq:ab_mod}
\end{align}
One can easily check that $a_{\kappa}^{kl} + b_{\kappa}^{kl} = 1$ and $\overline{\delta}_{\Lambda_c}^{kl}(V_LV_L) = 0$ are satisfied. 
Since the deviation vanishes in the SM limit, $\overline{\delta}_{\Lambda_c}^{kl}$ consists of NP contributions. 
Also, the sum rule \eqref{eq:HQ_sum_rule} is reproduced, {\it i.e.,} the left-hand side of Eq.~\eqref{eq:delta_mod_2} becomes zero, if we take the heavy quark limit, since $a_{\kappa}^{kl} \to a_{\rm HQL}$ and $b_{\kappa}^{kl} \to b_{\rm HQL}$ are satisfied in the limit.

\begin{figure}[!t]
\begin{center}
\includegraphics[width=0.45\linewidth]{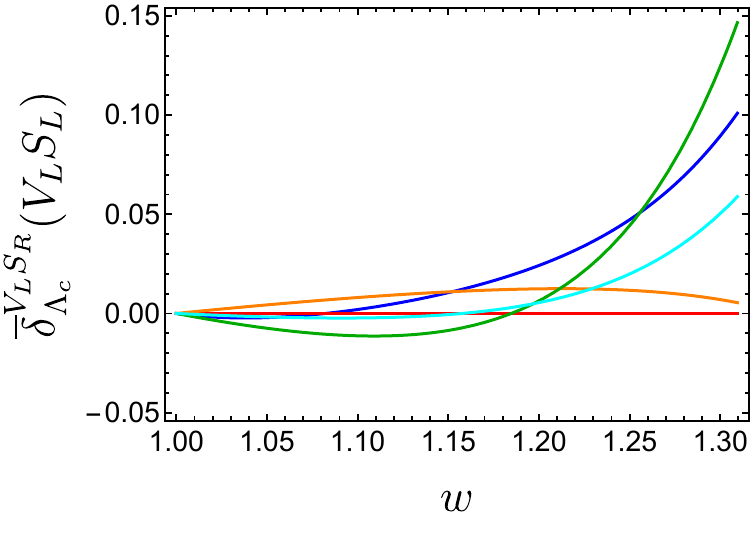}~~~
\includegraphics[width=0.45\linewidth]{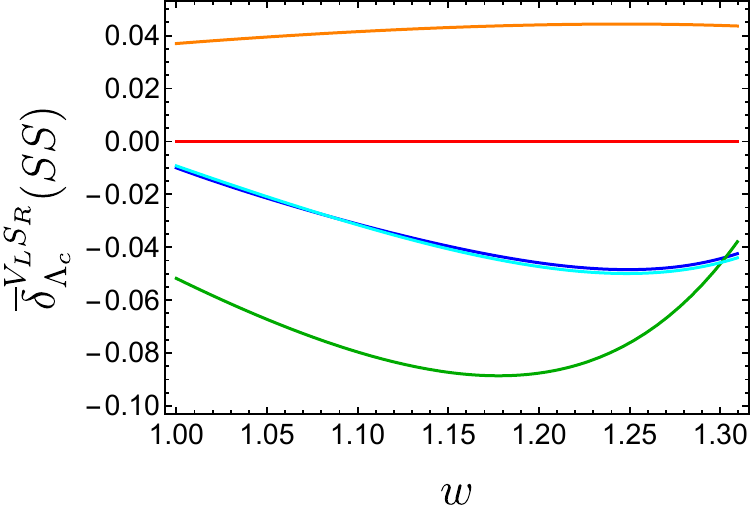} \vspace{0.2cm}\\
\includegraphics[width=0.45\linewidth]{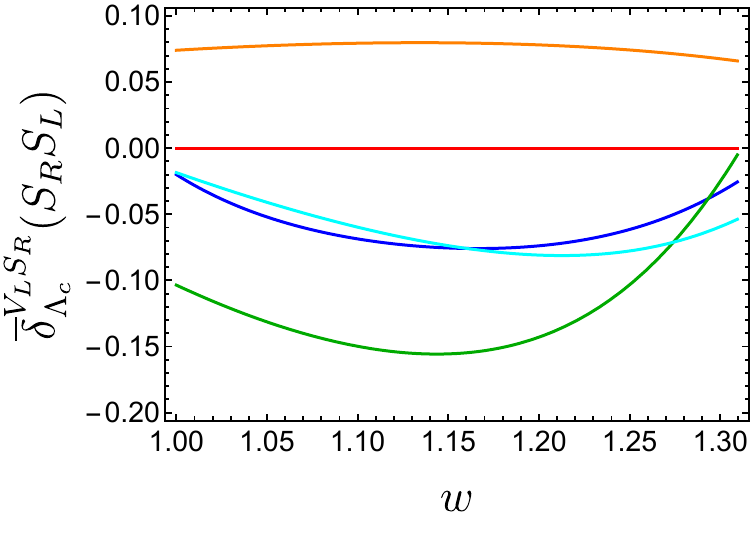}~~~~~~
\includegraphics[width=0.43\linewidth]{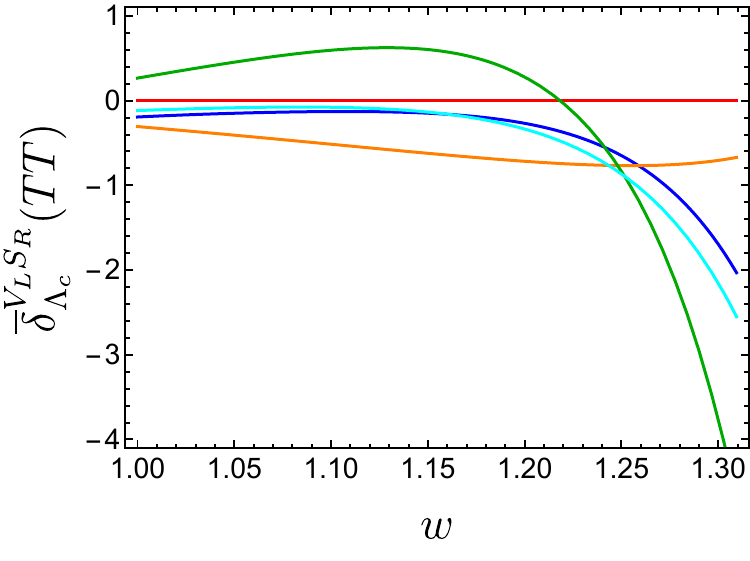} \vspace{0.2cm}\\
\includegraphics[width=0.45\linewidth]{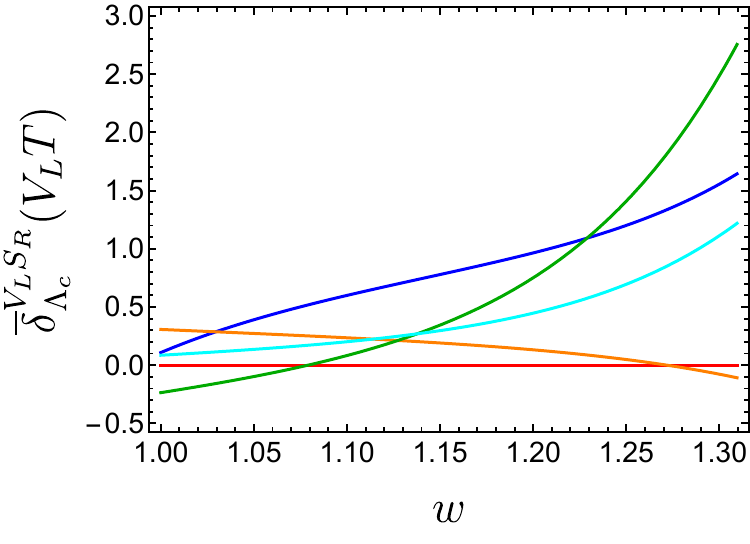}~~~~~~~~~~~
\mbox{\raisebox{12mm}{\includegraphics[width=0.37\linewidth]{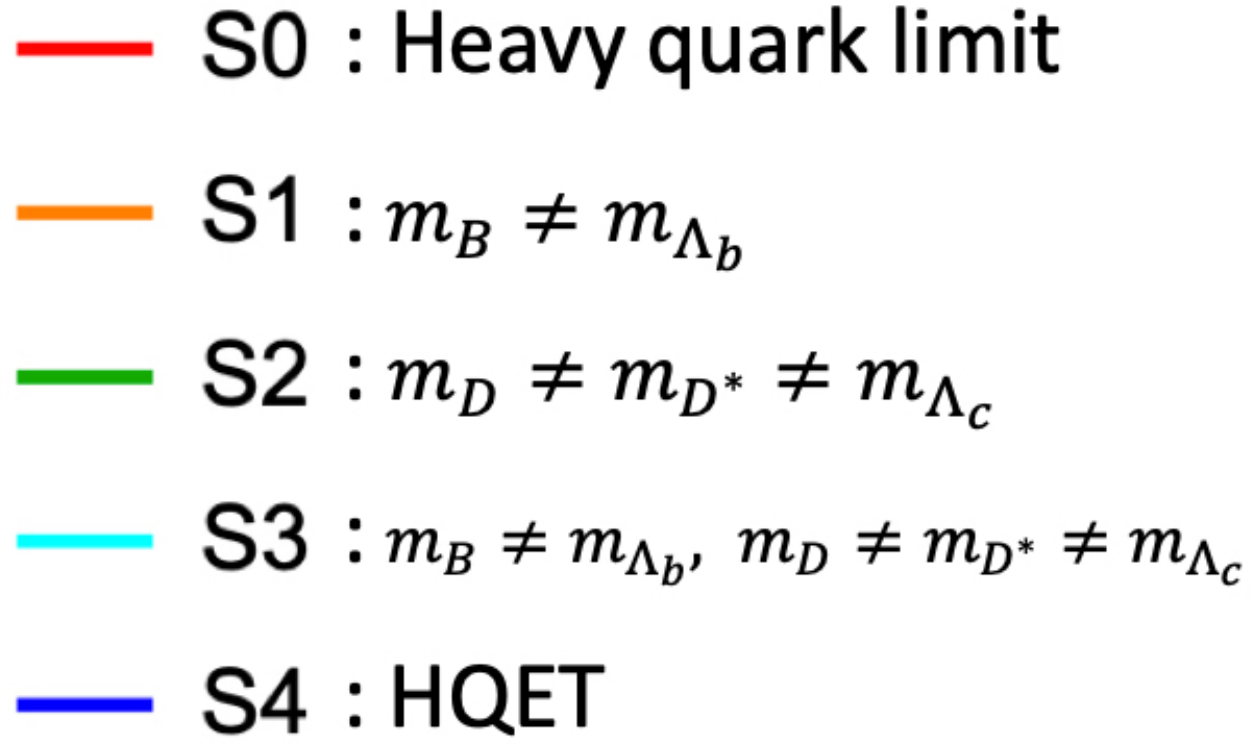}}}
\end{center}
\vspace{-.25cm}
\caption{
The deviation from the sum rule in the heavy quark limit, $\overline{\delta}^{V_LS_R}_{\Lambda_c}(ij)$, as a function of $w$. 
For further explanations of scenarios S0 -- S4, see the text.
}
\label{fig:dkappaVSR}
\end{figure}

As mentioned above, the sum rule \eqref{eq:HQ_sum_rule} is violated because the realistic bottomed and charmed mass spectra include the corrections \eqref{eq:mass_HQET}, and the form factors involve higher order contributions \eqref{eq:deltaFF}. 
For evaluating their effects on $\overline{\delta}_{\Lambda_c}^{kl}(ij)$, let us consider the following step-by-step scenarios: 
\begin{itemize}
    \item {S0: Heavy quark limit, {\it i.e.,} the scenario in Sec.~\ref{sec:SRinHQL}.}
    \item {S1: The form factors and the charmed hadron masses are the same as S0, but the bottomed hadrons have realistic mass spectra, {\it i.e.,} $m_B \neq m_{\Lambda_b}$.}
    \item {S2: The form factors and the bottomed hadron masses are the same as S0, but the charmed hadrons have realistic mass spectra, {\it i.e.,} $m_{D} \neq m_{D^*} \neq m_{\Lambda_c}$.}
    \item {S3: The form factors are the same as S0, but with realistic bottomed and charmed hadron mass spectra, $m_B \neq m_{\Lambda_b}$ and $m_{D} \neq m_{D^*} \neq m_{\Lambda_c}$.}
    \item {S4: The form factors include higher order corrections with realistic bottomed and charmed hadron mass spectra, $m_B \neq m_{\Lambda_b}$ and $m_{D} \neq m_{D^*} \neq m_{\Lambda_c}$.}
\end{itemize}
Scenario S4 corresponds to the full setup of the HQET. 

As a reference, let us consider a case when the $V_LS_R$ term is suppressed in Eq.~\eqref{eq:delta_mod_2}, {\it i.e.,} $\overline{\delta}^{V_LS_R}_{\Lambda_c}(V_LS_R) = 0$ is satisfied.\footnote{In Fig.~\ref{fig:app_1} of Appendix \ref{sec:add_fig} we show the results for other cases such as $\overline{\delta}^{SS}_{\Lambda_c}(ij)$ and $\overline{\delta}^{V_LS_R}_{\Lambda_c}(ij)$.}
In Fig.~\ref{fig:dkappaVSR}, $\overline{\delta}^{V_LS_R}_{\Lambda_c}(ij)$ is shown for $(ij)=(V_LS_L)$, $(SS)$, $(S_RS_L)$, $(TT)$, and $(V_LT)$ as functions of $w$.
In the heavy quark relation, the masses are commonly set to $m_D$ for the charmed hadrons in scenario S1, and $m_B$ for the bottomed hadrons in S2.
Also, the range of $w$ is taken from $1$ to $w_{\Lambda_c,\,max}$, since the kinematic phase space is not universal as $w_{\Lambda_c,\,max}<w_{D^*,\,max}<w_{D,\,max}$ for the realistic hadron masses.

From the figures, it is seen that $\overline{\delta}^{V_LS_R}_{\Lambda_c}(ij)=0$ is satisfied for S0 in the whole $w$ range (red line in the plot) because the scenario corresponds to the heavy quark limit. 
This exact relation is violated in S1 -- S4. 
Although S1 (orange) and S2 (green) predict deviations from zero, those contributions tend to be destructive to each other. 
Thus, S3 (cyan) provides a milder deviation than S1 and S2.
On the other hand, higher order contributions to the form factors in S4 (blue) also affect the sum rule. 
Their effects are comparable to those from the hadron mass spectra and could dominate the total deviations.

We also studied the $w$ dependence of the deviations.
They are relatively suppressed for small $w$. 
However, the result is based on the HQET and may depend on the form factor parametrization.
The analysis with the BGL form factors~\cite{Boyd:1995cf} will be explored in the following subsection. 
Also, we have not evaluated uncertainties of $\overline{\delta}^{kl}_{\Lambda_c}(ij)$, which will be studied elsewhere~\cite{Iguro:2025WIP}. 

The deviations for $(ij)=(TT)$ and $(V_LT)$ are larger than the others.
Such a tendency has also been seen in the $b \to c$ semileptonic sum rule \eqref{eq:R_sum_rule} (see Refs.~\cite{Blanke:2018yud, Blanke:2019qrx, Fedele:2022iib, Duan:2024ayo}).
Since the tensor contributions are likely to involve large uncertainties (see Ref.~\cite{Duan:2024ayo}), it is significant to accurately determine the form factors to obtain the sum rule reliably.

\subsection{Phase space integral}
\label{sec:integral}

In addition to the effects investigated in the previous subsection, the equality of the sum rule in the heavy quark limit \eqref{eq:HQ_sum_rule} is violated by phase space integrals of the differential decay rates. 
Experimentally, the decay rates are measured in the whole range/intervals of $w$. 
Additionally, the sum rule was provided originally for the total decay rates (see Eq.~\eqref{eq:R_sum_rule}). 
Then, in each term of Eq.~\eqref{eq:delta_mod_2}, the ratios are replaced as
\begin{align}
 \frac{\kappa^{ij}_{H_c}}{\kappa_{H_c}^{V_LV_L}} \to \frac{\gamma^{ij}_{H_c,I}}{\gamma^{V_LV_L}_{H_c,I}} \,,~~~~~
 \gamma^{ij}_{H_c,I} \equiv \int_{w_1}^{w_2}\!\kappa^{ij}_{H_c}\,dw \,,
 \label{eq:kappa_mod}
\end{align}
where $I = [w_1,w_2]$ is the integral interval.
Instead of $\overline{\delta}_{\Lambda_c}^{kl}$ in Eq.~\eqref{eq:delta_mod_2}, we consider the following deviation,
\begin{align}
 \delta_{\Lambda_c}^{kl} = \sum_{ij}  \mathcal{C}_i \mathcal{C}_j^* \,\delta_{\Lambda_c}^{kl}(ij)\,,
 ~~~
 \delta_{\Lambda_c}^{kl}(ij) \equiv 
 \frac{\gamma^{ij}_{\Lambda_c,I}}{\gamma^{V_LV_L}_{\Lambda_c,I}} - a^{kl}_I \frac{\gamma^{ij}_{D,I}}{\gamma^{V_LV_L}_{D,I}} - b^{kl}_I \frac{\gamma^{ij}_{D^*,I}}{\gamma^{V_LV_L}_{D^*,I}} \,.
 \label{eq:delta_mod_int}
\end{align}
The coefficients $a^{kl}_I$ and $b^{kl}_I$ are no longer functions of $w$ and given by
\begin{align}
 a^{kl}_I = \frac{\gamma_{\Lambda_c,I}^{kl}/\gamma_{\Lambda_c,I}^{V_LV_L}-\gamma_{D^*,I}^{kl}/\gamma_{D^*,I}^{V_LV_L}}{\gamma_{D,I}^{kl}/\gamma_{D,I}^{V_LV_L}-\gamma_{D^*,I}^{kl}/\gamma_{D^*,I}^{V_LV_L}} \,,~~~
 b^{kl}_I = \frac{\gamma_{D,I}^{kl}/\gamma_{D,I}^{V_LV_L}-\gamma_{\Lambda_c,I}^{kl}/\gamma_{\Lambda_c,I}^{V_LV_L}}{\gamma_{D,I}^{kl}/\gamma_{D,I}^{V_LV_L}-\gamma_{D^*,I}^{kl}/\gamma_{D^*,I}^{V_LV_L}} \,.
 \label{eq:ab_mod_int}
\end{align}
Here, $a^{kl}_I + b^{kl}_I = 1$, $\delta_{\Lambda_c}^{kl}(kl) = 0$, and $\delta_{\Lambda_c}^{kl}(V_LV_L) = 0$ are satisfied. 
In particular, the sum rule is exact in the SM limit.

\begin{figure}[!t]
\begin{center}
\includegraphics[width=0.45\linewidth]{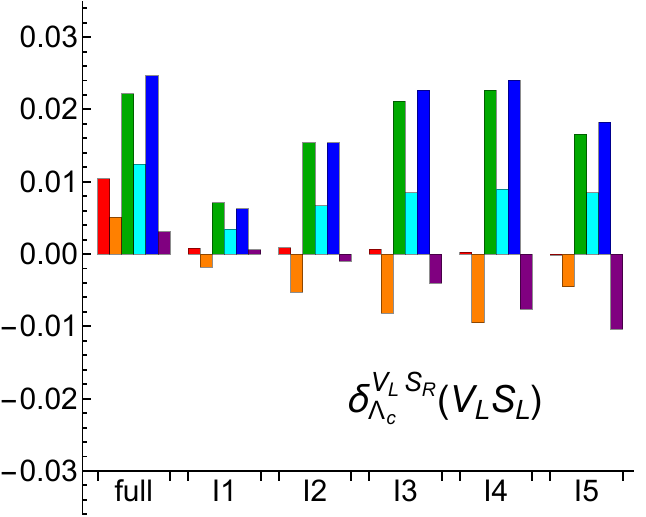}~
\includegraphics[width=0.45\linewidth]{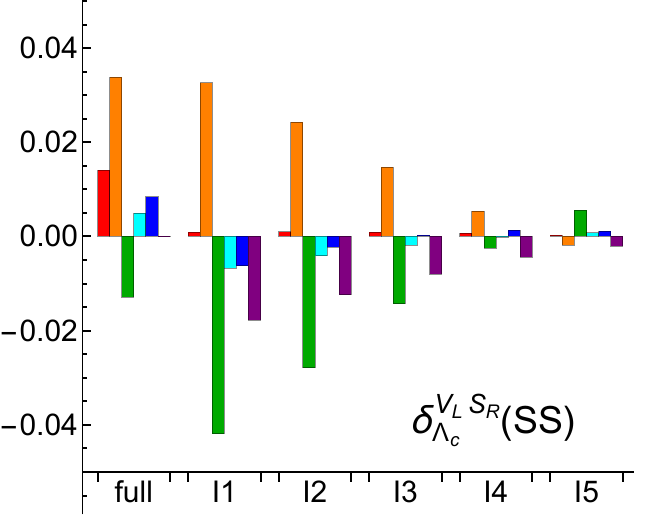}\\\vspace{0.3cm}
\includegraphics[width=0.45\linewidth]{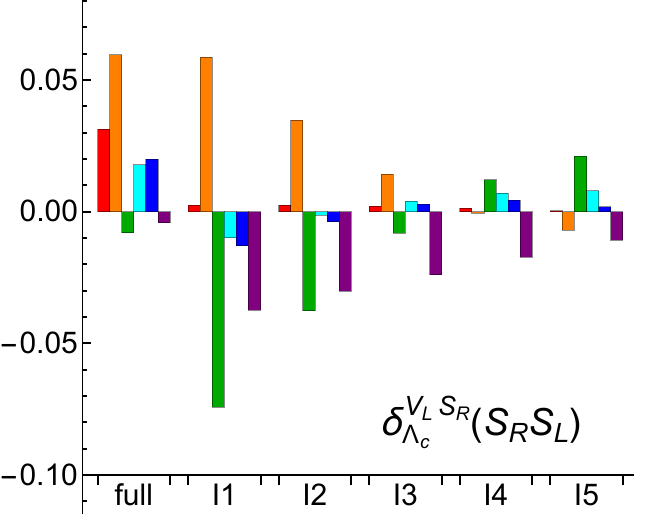}~
\includegraphics[width=0.45\linewidth]{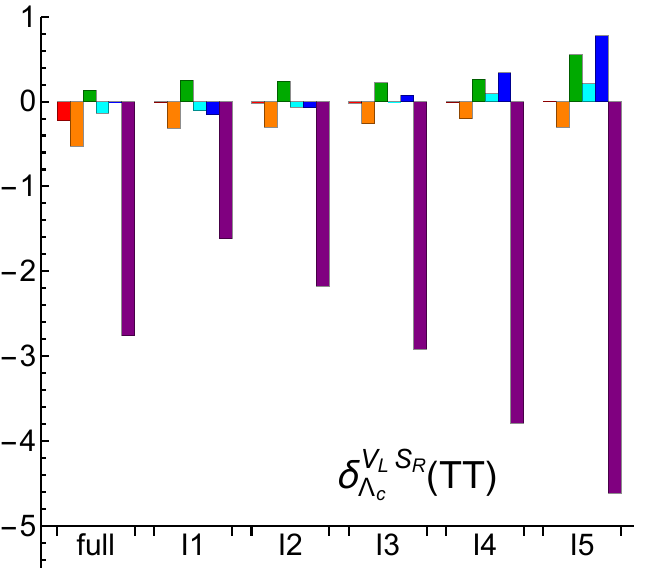}\\\vspace{0.3cm}
\includegraphics[width=0.45\linewidth]{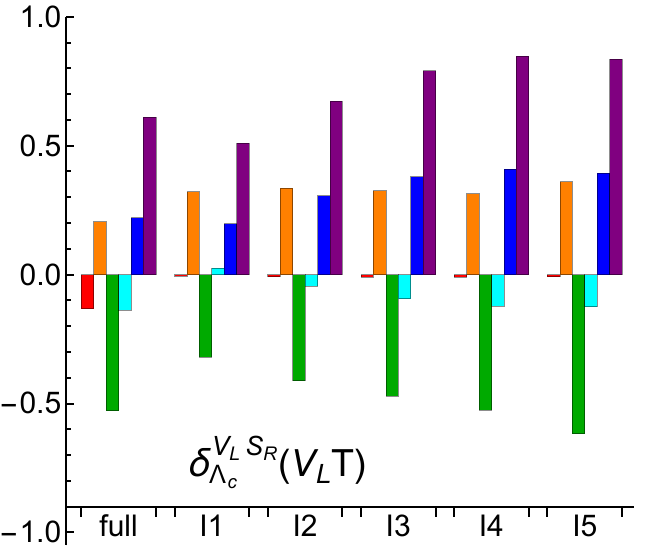}
~~~~~~~~
\mbox{\raisebox{10mm}{\includegraphics[width=0.35\linewidth]{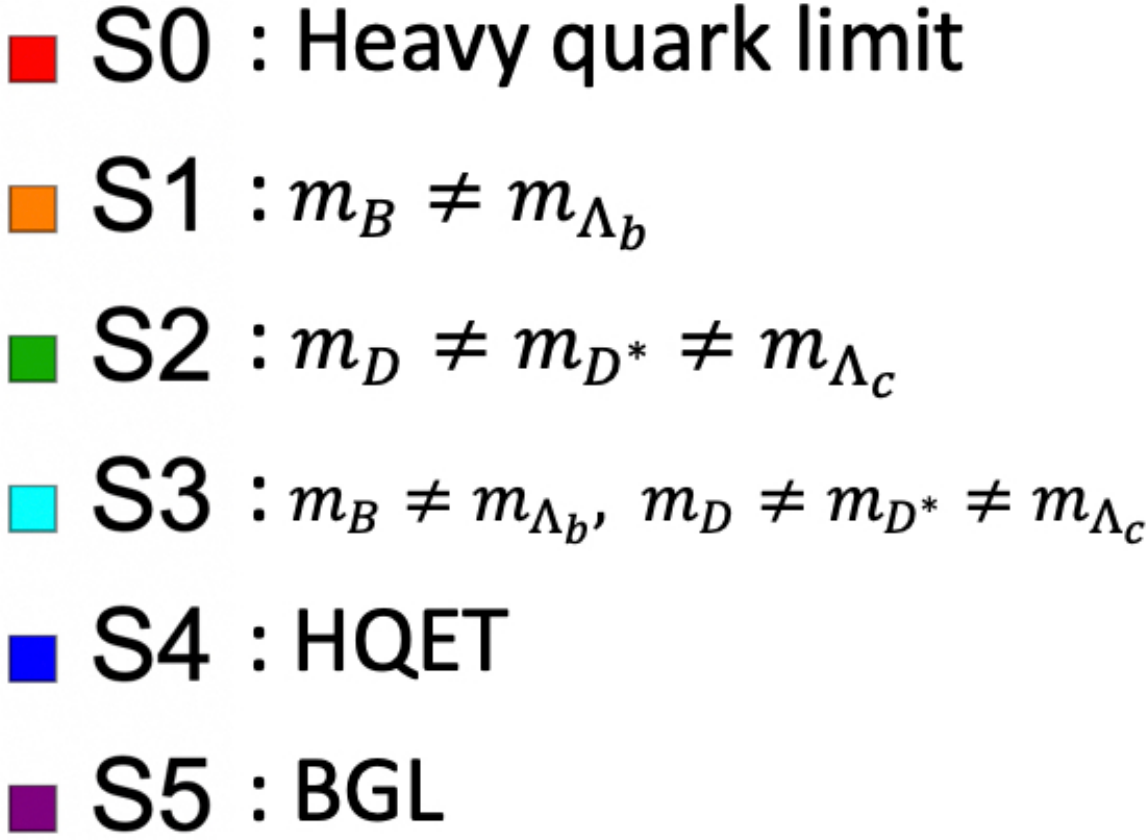}}}
\end{center}
\vspace{-.25cm}
\caption{The correction to the sum rule, $\delta_{\Lambda_c}^{V_LS_R}(ij)$, for five $w$ intervals and the total decay rates (``full'').
Different colors correspond to different scenarios also explained in the main text.}
\label{fig:dintVSR}
\end{figure}

For the integral intervals, we consider the following five $w$ regions,
\begin{align}
 Ii = [1+(w_{H_c,\,max}-1)(i-1)/5,\,\,1+(w_{H_c,\,max}-1)i/5] \,.~~~
 (i = 1, \dots, 5)
 \label{eq:w_interval}
\end{align}
The region is commonly set between the numerator and denominator of each ratio in Eqs.~\eqref{eq:delta_mod_int} and \eqref{eq:ab_mod_int}. 
We also study the case of the total decay rates, {\it i.e.,} the one that the phase space is integrated over $1 < w < w_{H_c,\,max}$, corresponding to the $b \to c$ semileptonic sum rule \eqref{eq:R_sum_rule}.

Once the phase space is integrated, unlike the previous sections, the leading order IW functions do not cancel in each ratio of Eqs.~\eqref{eq:delta_mod_int} and \eqref{eq:ab_mod_int}, necessitating their explicit $w$ dependence.
We assume the functions as~\cite{Bernlochner:2017jka, Bernlochner:2018bfn}\footnote{
In detail, $\xi^{\rm{pre}}(w)$ is used only for the leading order evaluations of the form factors in the following. 
For S4, we adopt a different parameterization explored in Ref.~\cite{Iguro:2020cpg}. } 
\begin{align}
  &\xi(w) = \xi^{\rm{pre}}(w)/\xi^{\rm{pre}}(1) \,,~~~~
   \xi^{\rm{pre}}(w) = 1 - 8a^2\overline{\rho^2_*}z_* + \left( V_{21}\overline{\rho^2_*}-V_{20}\right) z_*^2 \,, \\
  &\zeta(w) = 1 + (w-1) \zeta^\prime + \frac{1}{2}(w-1)^2 \zeta^{\prime\prime} \,,
\end{align}
where $z_*$ is a function of $w$,
\begin{align}
 z_* = \frac{\sqrt{w+1}-\sqrt{2}a}{\sqrt{w+1}+\sqrt{2}a} \,.
\end{align}
Here, $a^2\simeq1.14$, $V_{21}\simeq57$, $V_{20}\simeq7.5$, and $\overline{\rho^2_*} \simeq  1.24$ are introduced for $\xi(w)$~\cite{Bernlochner:2017jka}\footnote{In the literature, the value of $\overline{\rho^2_*}$ is determined along with other sub-leading IW parameters by using Belle data, Lattice, and QCDSR. 
However, they are ignored in our analysis for simplicity.
}.
Also, $\zeta^\prime \simeq -2.06$ and $\zeta^{\prime\prime} \simeq 3.28$ are set for $\zeta(w)$~\cite{Bernlochner:2018bfn}.\footnote{These values are determined by performing a global fit to LHCb data~\cite{LHCb:2017vhq} and Lattice result~\cite{Detmold:2015aaa}, where the analyses are based on the realistic hadron mass spectra. 
}

So far, we have focused on the HQET. 
In the literature, the sum rule has been explored with BGL form factors as well as those in the HQET (see Refs~\cite{Blanke:2018yud, Blanke:2019qrx, Fedele:2022iib, Duan:2024ayo}).\footnote{
In the literature, the $\Lambda_b\to\Lambda_c$ transitions have been studied only with the BGL form factors for the sum rule. 
}
In the following, we evaluate $\delta_{\Lambda_c}^{kl}(ij)$ based on the BGL form factors by following Ref.~\cite{Duan:2024ayo}. 
Hence, in addition to S0 -- S4, we consider scenario S5,
\begin{itemize}
    \item {S5: the BGL form factors with realistic hadron masses.\footnote{
    We do not evaluate uncertainties. 
    Further study is in progress~\cite{Iguro:2025WIP}.} 
    }
\end{itemize}

In Fig.~\ref{fig:dintVSR}, we show $\delta_{\Lambda_c}^{V_LS_R}(ij)$ for the case when the differential decay rates are integrated over $w$ in the range of I1 -- I5 and the case for the total decay rate (denoted as ``full'').\footnote{
 Other results such as $\delta^{V_LS_L}_{\Lambda_c}(ij)$ and $\delta^{V_LT}_{\Lambda_c}(ij)$ are given in Fig.~\ref{fig:app_2} of Appendix \ref{sec:add_fig}.
}
The phase space integral makes the deviation $\delta_{\Lambda_c}^{V_LS_R}(ij)$ non-zero even in the heavy quark limit (S0, red bar in the plot).
The violations of the heavy quark relation in the hadron mass spectra in S1 (orange) and S2 (green) induce large deviations similar to Fig.~\ref{fig:dkappaVSR}.
These two contributions tend to be destructive to each other, and thus, the deviation in S3 (cyan) becomes milder.
The deviations from the higher order contributions to the form factors (S4, blue) are comparable to or larger than those in S3. 
On the other hand, the BGL form factors (S5, purple) predict smaller deviations than those in the HQET (S4) except for $(ij) = (TT)$ and $(V_LT)$. 
In particular, for $(ij)=(V_LS_L)$ the former provides a much smaller deviation than the latter in the case of the total decay rate (``full''). 
This might be interesting because NP contributions to $C_{S_L}$ are favored by the $R_{D^{(*)}}$ anomaly.\footnote{
See Refs.~\cite{Iguro:2022uzz,Blanke:2022pjy,Iguro:2023jju,Crivellin:2023sig,Athron:2024rir} for recent works on the light $H^\pm$ contribution to $C_{S_L}$ under stringent collider constraints of Ref.~\cite{Iguro:2018fni}, though evaluations of $\delta_{\Lambda_c}^{V_LS_R}(V_LS_L)$ uncertainties are left for future works. 
}
For $(ij) = (TT)$ and $(V_LT)$, the corrections to the sum rule seem to deviate from zero significantly in S5.
This means that when we apply the sum rule to check the consistency in the experimental results, we may be unable to ignore the corrections as long as the NP would appear largely in the tensor operators.
However, a typical size of $C_T$ is $\mathcal{O}(10^{-2})$ from the current $R_D$ and $R_{D^*}$ anomalies, and also, it has been constrained by the LHC results as $\lesssim 0.1-0.3$ depending on the setup~\cite{Iguro:2024hyk}. 
Thus, the corrections to the sum rule turn out to be minor (see Ref.~\cite{Duan:2024ayo}), in particular, compared to the experimental uncertainty of $R_{\Lambda_c}$~\cite{LHCb:2022piu, Bernlochner:2022hyz}.
Nonetheless, the large differences between S4 and S5 indicate that there should be sizable (potential) uncertainties in the form factors, which are required to be determined precisely.

As for the $w$ dependence, among the five $w$ intervals I1 -- I5, I1 is likely to provide smaller deviations in the HQET (S4), as expected from Fig.~\ref{fig:dkappaVSR}. 
However, such a tendency is not seen in S5, {\it i.e.,} with the BGL form factors.
Therefore, it is inevitable to determine the form factors more accurately for further discussion.

\section{Summary and discussion}
\label{sec:Summary}

The $b\to c$ semileptonic sum rule \eqref{eq:R_sum_rule} has attracted attention in light of the $R_D$ and $R_{D^*}$ anomalies. 
Although the rule involves a model-dependent correction $\delta_{\Lambda_c}$, its smallness enables us to apply the relation to check the consistency in the experimental results independently of NP models. 

Despite its usefulness, the sum rule was found empirically, and its theoretical background was not clear. 
In this paper, we investigated $\bar B\to D\tau\bar\nu_\tau$ and $\bar B\to D^*\tau\bar\nu_\tau$, and   $\Lambda_b\to \Lambda_c\tau\bar\nu_\tau$ from the perspective of the heavy quark symmetry. 
Among their differential decay rates, we derived a sum rule holding exactly in the heavy quark limit, {\it i.e.,} there is no model-dependent correction. 

We then analyzed how the equality of the sum rule is violated in reality. 
The corrections are introduced by the heavy quark symmetry breaking of the form factors and the violation of the heavy quark relation in the hadron mass spectra.
Also, the phase space integrals of the differential decay rates affect the sum rule even in the heavy quark limit. 
We found a large cancellation between the contributions from the bottomed and charmed hadron mass spectra. 
Additionally, the higher order contributions to the form factors yield a deviation comparable to or larger than those from the hadron mass spectra. 
Therefore, the accuracy of the form factors is crucial for further investigations of the sum rule. 

We also compared the results obtained in the HQET with those based on the BGL form factors. 
It was shown that there are large differences between these two approaches, indicating (potential) uncertainties associated with the form factors and other inputs, though they are not evaluated in this paper.  
Those arising from parameterizations, alongside investigating the phenomenological impact based on the best-fit values from Ref.~\cite{Iguro:2024hyk} and their $w$ dependencies, remain a subject for future work.

\section*{Acknowledgements}
The authors thank Ulrich Nierste, Martin S. Lang, Teppei Kitahara, Hiroyasu Yonaha, Junji Hisano, and Takashi Kaneko for the inspiring and encouraging discussions.
This work is supported by JSPS KAKENHI Grant Numbers 21H01086 [M.E.], 22K21347 [M.E. and S.I.], 23K20847 [M.E.], 24K07025 [S.M.], and 24K22879 [S.I.].
The work of S.I. is also supported by JPJSCCA20200002 and the Toyoaki scholarship foundation.
S.I. appreciates the joint appointment program between Nagoya University and KEK which allowed him to start this project. 
We thank KMI for the visitor program that allowed S.I. to invite R.W. and have on-site discussions at Nagoya University.

\appendix
\section{Hadronic amplitudes}
\label{sec:helicity}

The hadronic factors of the helicity amplitudes in the HQET are summarized in this appendix. 
For the $B \to D$ transitions, there are four amplitudes~\cite{Iguro:2020cpg, Sakaki:2013bfa}:
\begin{align}
 H_{V,0}^D &= m_B \sqrt{\frac{r_D(w^2-1)}{\hat q_D^2}} \Big[ (1+r_D)h_+ -(1-r_D)h_- \Big] \,, \notag \\
 H_{V,t}^D &= m_B \sqrt{\frac{r_D}{\hat q_D^2}} \Big[ (1-r_D)(w+1)h_+ -(1+r_D)(w-1)h_- \Big] \,, \notag \\
 H_{S}^D &= m_B \sqrt{r_D} (w+1) h_S \,, \notag \\
 H_{T}^D &= - m_B \sqrt{r_D(w^2-1)}\, h_T\,.
\end{align}
For the $B \to D^*$ transitions, they are given by~\cite{Iguro:2020cpg, Sakaki:2013bfa}
\begin{align}
 H_{V,\pm}^{D^*} 
 &= m_B \sqrt{r_{D^*}} \Big[ (w+1) h_{A_1} \mp \sqrt{w^2-1} h_{V} \Big] \,, 
 \notag \\ 
 H_{V,0}^{D^*} 
 &= m_B \sqrt{\frac{r_{D^*}}{\hat q_{D^*}^2}} (w+1) 
 \Big[ (r_{D^*}-w) h_{A_1} + (w-1) (r_{D^*} h_{A_2} + h_{A_3}) \Big] \,, 
 \notag \\
 H_{V,t}^{D^*} 
 &= -m_B \sqrt{\frac{r_{D^*}(w^2-1)}{\hat q_{D^*}^2}}
 \Big[ (w+1) h_{A_1} + (r_{D^*}w-1) h_{A_2} + (r_{D^*}-w) h_{A_3} \Big] \,, 
 \notag \\
 H_{P}^{D^*} 
 &= - m_B \sqrt{r_{D^*}(w^2-1)} h_P \,, 
 \notag \\ 
 H_{T,\pm}^{D^*} 
 &= \pm m_B \sqrt{\frac{r_{D^*}}{\hat q_{D^*}^2}} \left[ 1-r_{D^*} (w \mp \sqrt{w^2-1}) \right]
 \left[ h_{T_1} + h_{T_2} + \left( w \pm \sqrt{w^2-1} \right) (h_{T_1} - h_{T_2}) \right] \,, 
 \notag \\
 H_{T,0}^{D^*} 
 &= -m_B \sqrt{r_{D^*}} \Big[ (w+1) h_{T_1} +(w-1) h_{T_2} +2(w^2-1) h_{T_3} \Big] \,.
\end{align}
For $\Lambda_b \to \Lambda_c$, the hadronic factors are shown as~\cite{Datta:2017aue, Bernlochner:2018bfn}
\begin{align}
 H_{V,1\pm}^\Lambda 
 &= -2 \,m_{\Lambda_b} \sqrt{r_\Lambda} \Big[ \sqrt{w-1} f_1 \mp \sqrt{w+1} g_1 \Big] \,,
 \notag \\
 H_{V,0\pm}^\Lambda 
 &= m_{\Lambda_b} \sqrt{\frac{2r_\Lambda}{\hat q_\Lambda^2}} 
 \Big\{
 \sqrt{w-1} \Big[(1+r_\Lambda)f_1 + (w+1)(f_2 r_\Lambda + f_3)\Big]
 \notag \\ 
 & \qquad\qquad\quad
 \mp \sqrt{w+1} \Big[(1-r_\Lambda)g_1 - (w-1)(g_2 r_\Lambda + g_3)\Big]
 \Big\} \,, 
 \notag \\
 H_{V,t\pm}^\Lambda 
 &= m_{\Lambda_b} \sqrt{\frac{2r_\Lambda}{\hat q_\Lambda^2}} \Big\{
 \sqrt{w+1} \Big[(1-r_\Lambda)f_1 + f_2(1-w r_\Lambda) + f_3(w-r_\Lambda)\Big]
 \notag \\ 
 & \qquad\qquad\quad
 \mp \sqrt{w-1} \Big[(1+r_\Lambda)g_1 - g_2(1-w r_\Lambda) - g_3(w-r_\Lambda)\Big] \Big\} \,, 
 \notag \\
 H_{S}^\Lambda 
 &= m_{\Lambda_b} \sqrt{2r_\Lambda(w+1)} h'_S \,, 
 \notag \\
 H_{P}^\Lambda 
 &= m_{\Lambda_b} \sqrt{2r_\Lambda(w-1)} h'_P \,, 
 \notag \\
 H_{T,1\pm}^\Lambda 
 &= -2 m_{\Lambda_b} \sqrt{\frac{r_\Lambda}{\hat q_\Lambda^2}} 
 \Big\{ 
 \sqrt{w-1} \Big[ (1+r_\Lambda) h_1 - (1-wr_\Lambda) h_2 - (w-r_\Lambda) h_3 \Big] 
 \notag \\ 
 & \qquad\qquad\qquad
 \pm \sqrt{w+1} \Big[ (1-r_\Lambda) h_1 - (w-1) (h_2 r_\Lambda + h_3) \Big] 
 \Big\} \,, 
 \notag \\
 H_{T,0\pm}^\Lambda 
 &= m_{\Lambda_b} \sqrt{2r_\Lambda} \Big\{ \sqrt{w-1} \Big[h_1-h_2+h_3-(w+1)h_4\Big] \pm \sqrt{w+1}h_1 \Big\} \,.
\end{align}

\section{Additional figures}
\label{sec:add_fig}

\begin{figure}[!t]
\begin{center}
\includegraphics[width=0.21\linewidth]{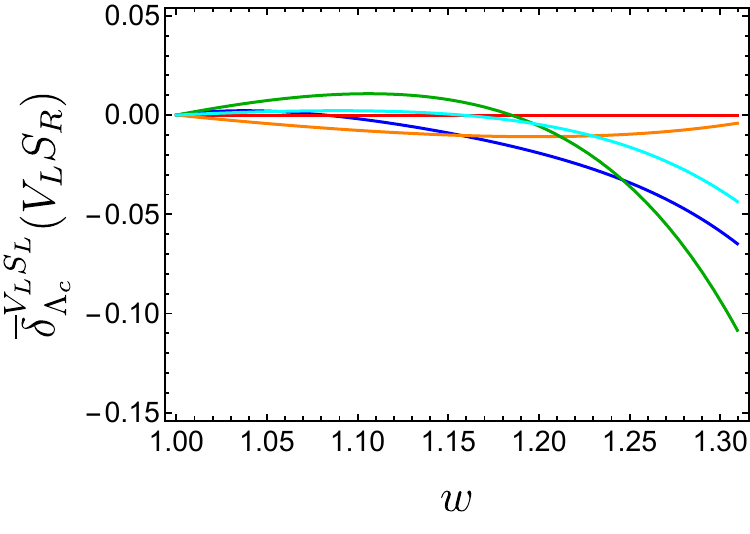}~
\includegraphics[width=0.21\linewidth]{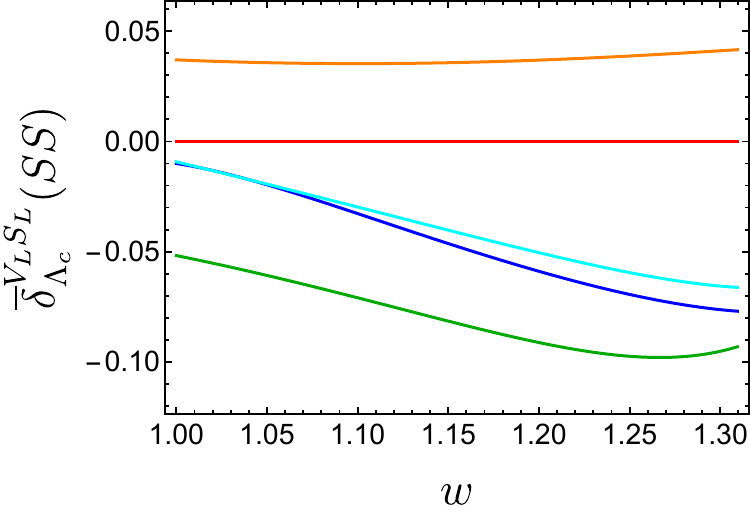}~
\includegraphics[width=0.21\linewidth]{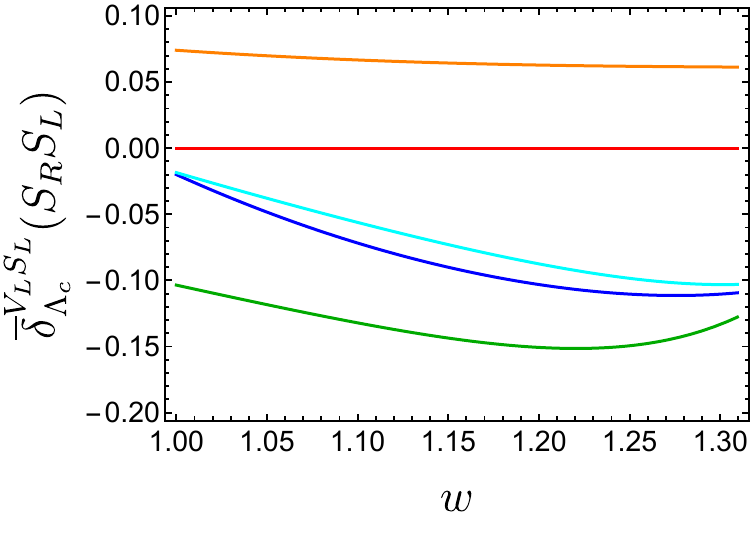}~
\includegraphics[width=0.21\linewidth]{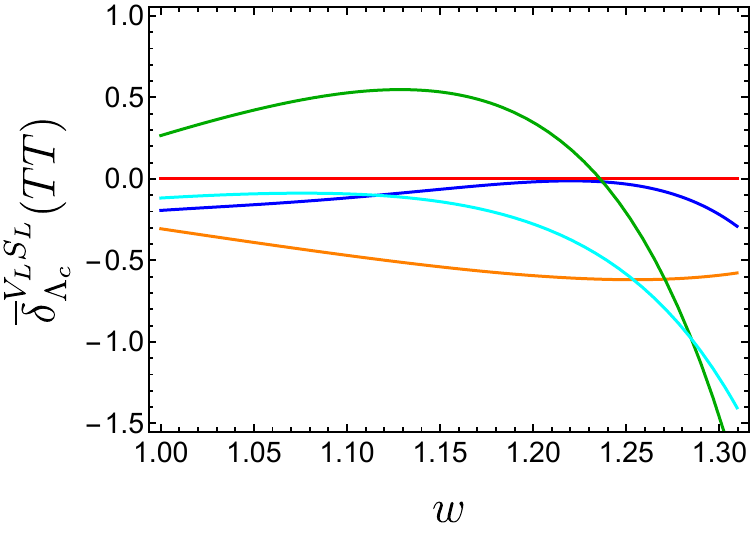}~
\includegraphics[width=0.21\linewidth]{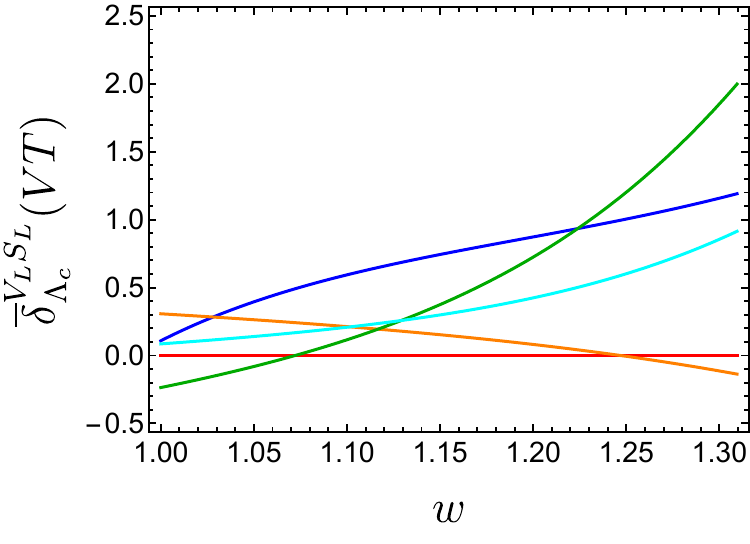}\\ \vspace{0.1cm}
\includegraphics[width=0.21\linewidth]{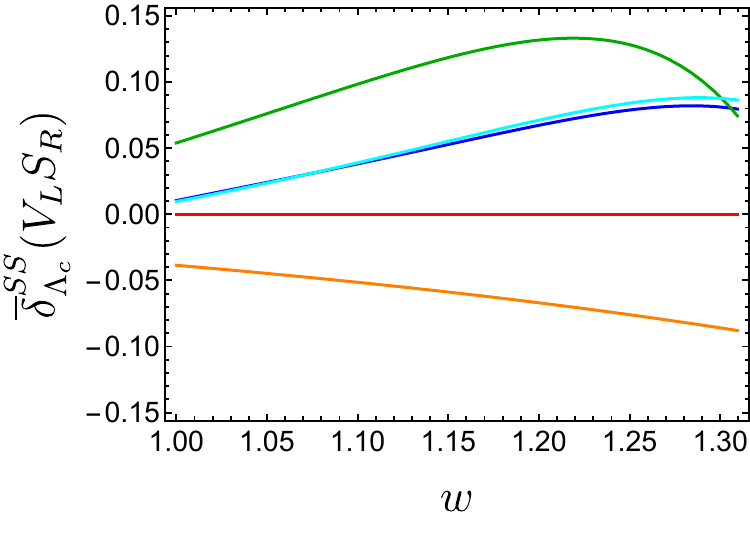}~
\includegraphics[width=0.21\linewidth]{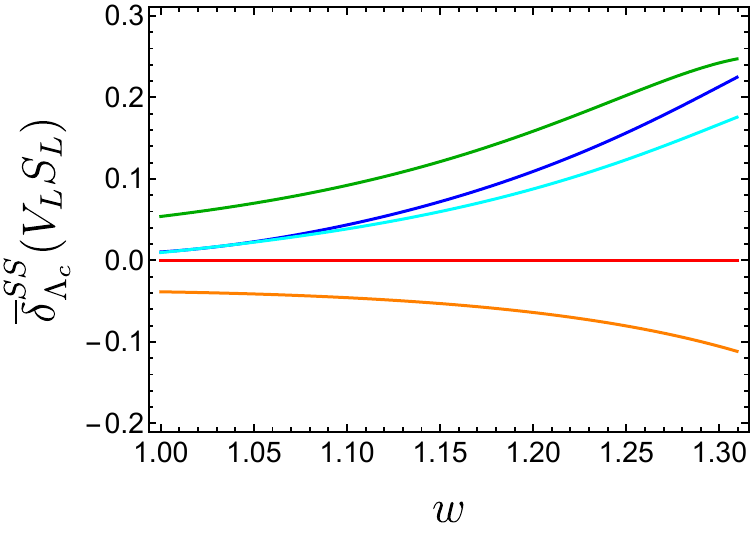}~
\includegraphics[width=0.21\linewidth]{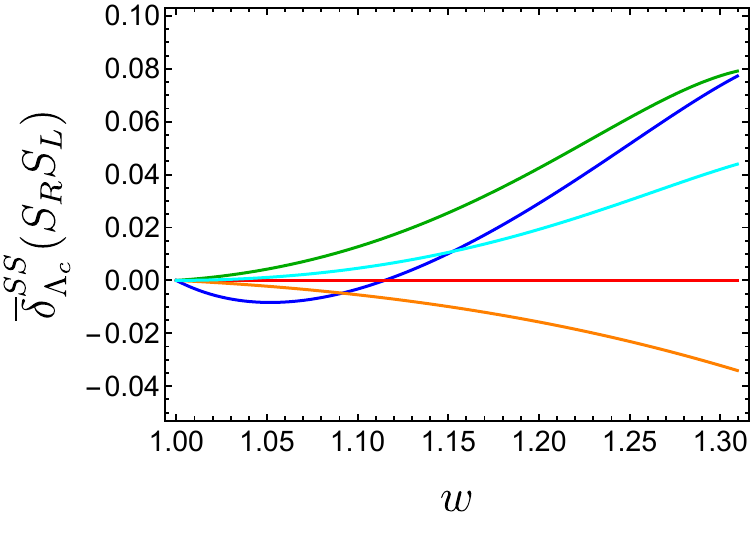}~
\includegraphics[width=0.21\linewidth]{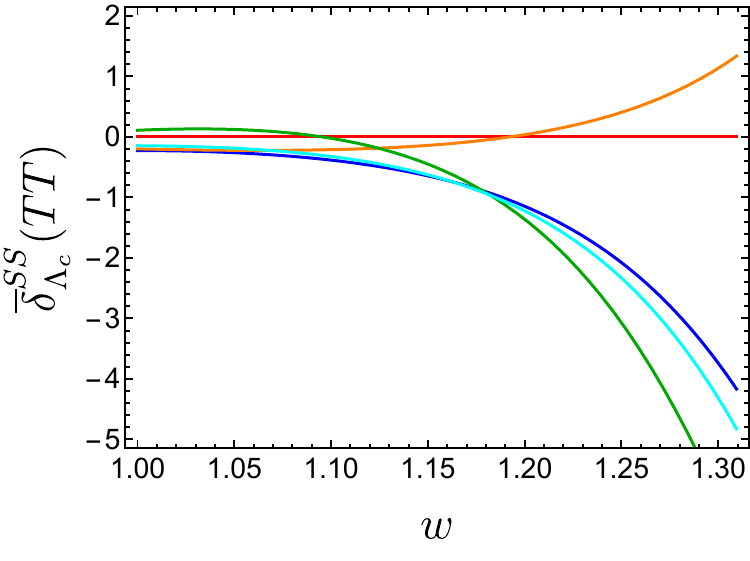}~
\includegraphics[width=0.21\linewidth]{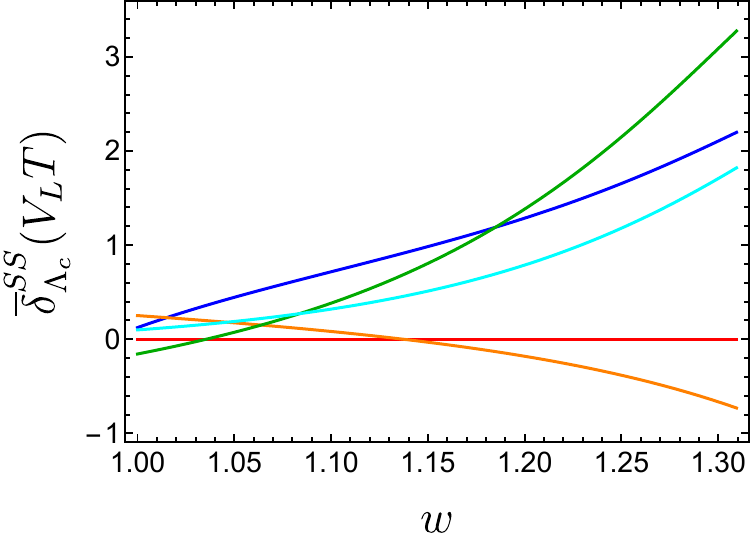}\\ \vspace{0.1cm}
\includegraphics[width=0.21\linewidth]{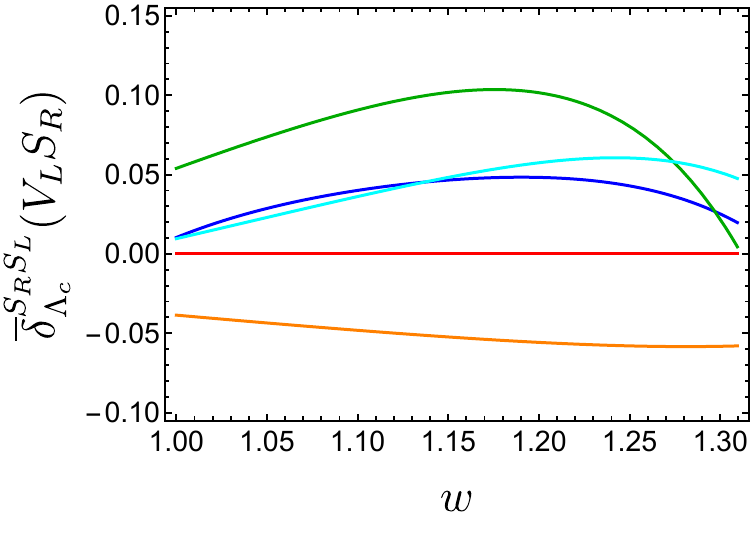}~
\includegraphics[width=0.21\linewidth]{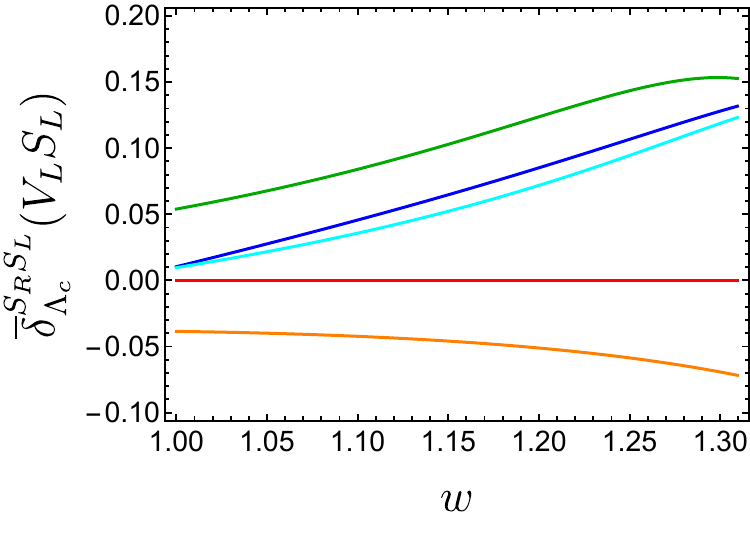}~
\includegraphics[width=0.21\linewidth]{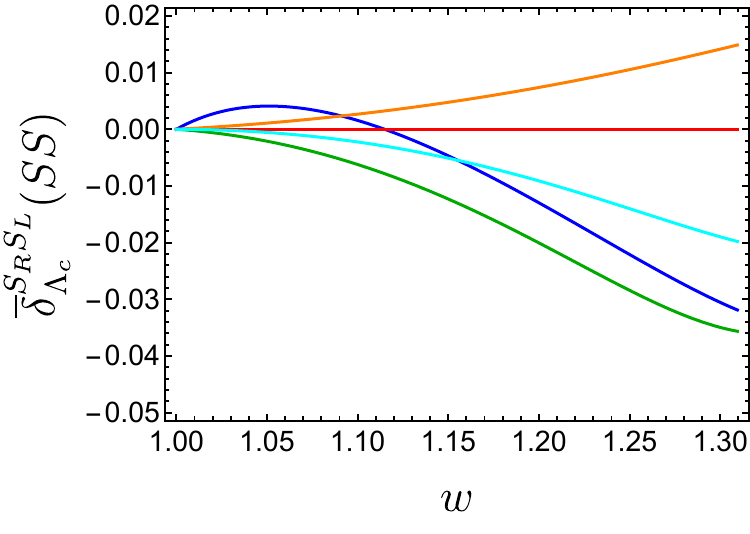}~
\includegraphics[width=0.21\linewidth]{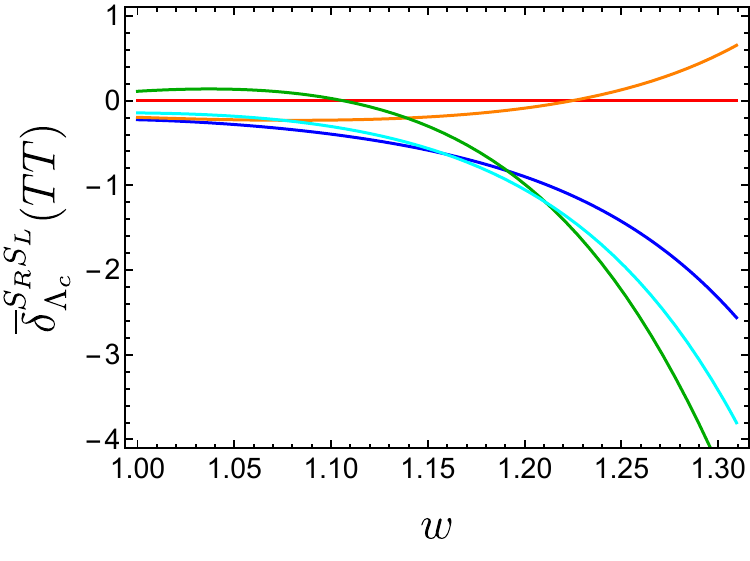}~
\includegraphics[width=0.21\linewidth]{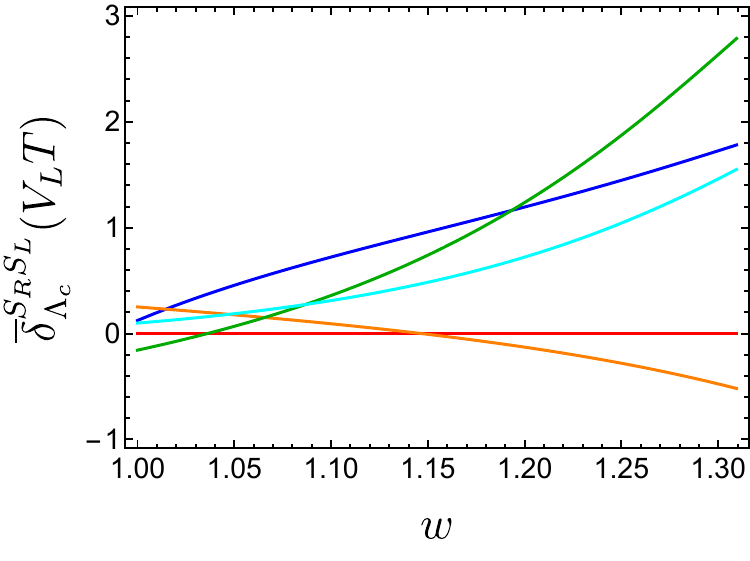}\\ \vspace{0.1cm}
\includegraphics[width=0.21\linewidth]{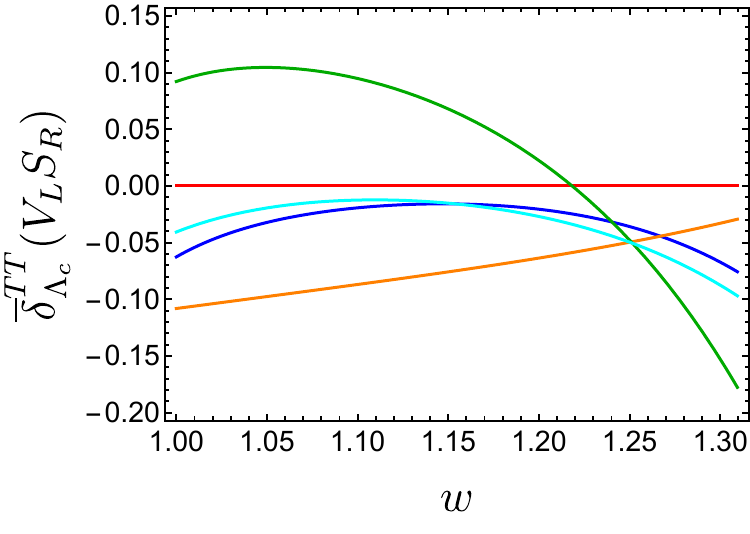}~
\includegraphics[width=0.21\linewidth]{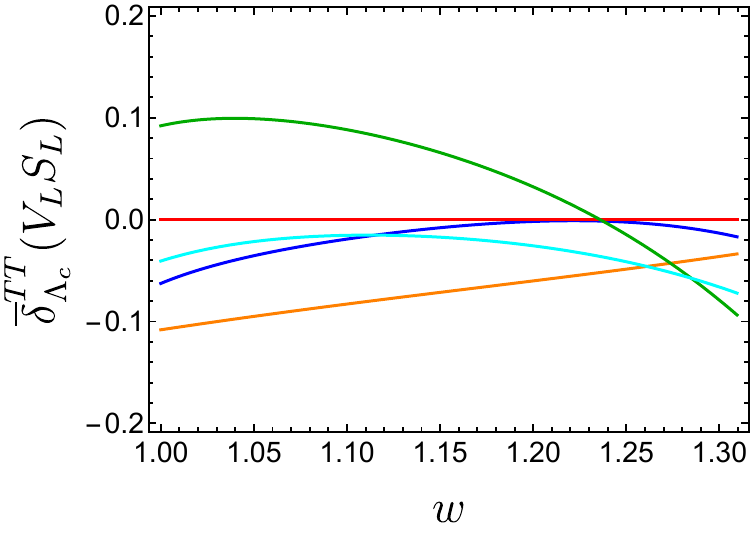}~
\includegraphics[width=0.21\linewidth]{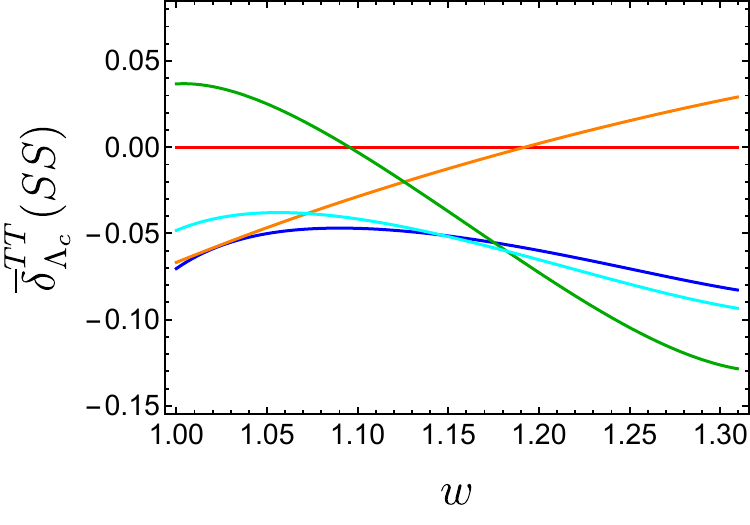}~
\includegraphics[width=0.21\linewidth]{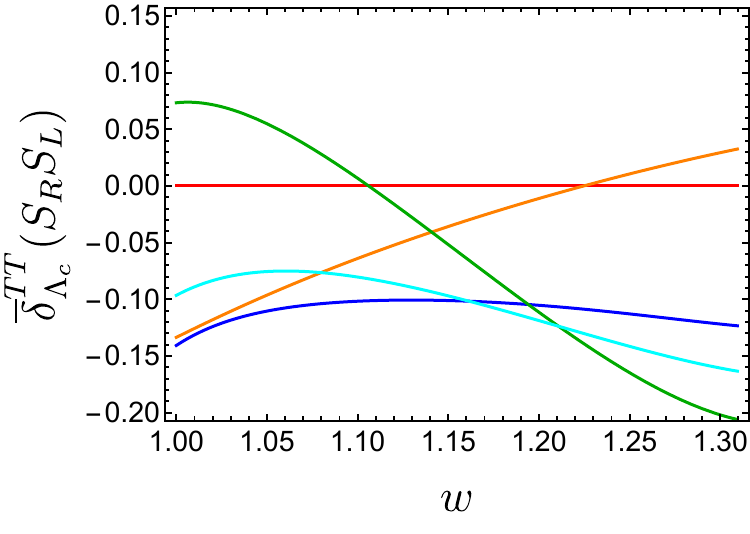}~
\includegraphics[width=0.21\linewidth]{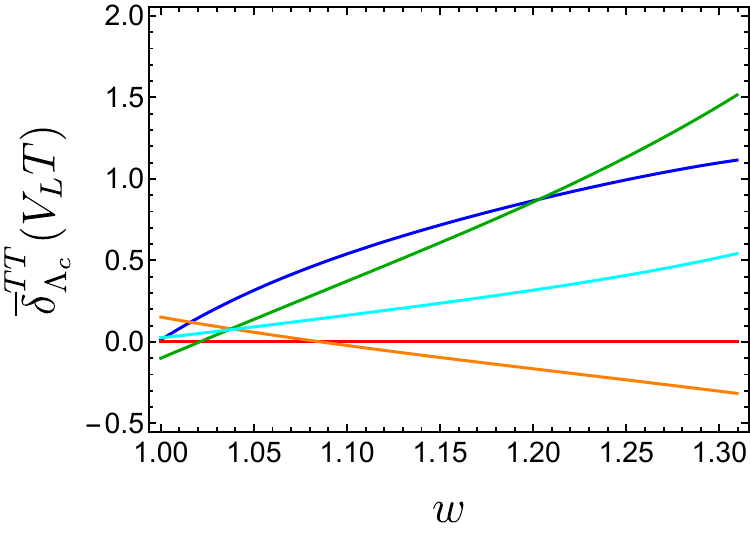}\\ \vspace{0.1cm}
\includegraphics[width=0.21\linewidth]{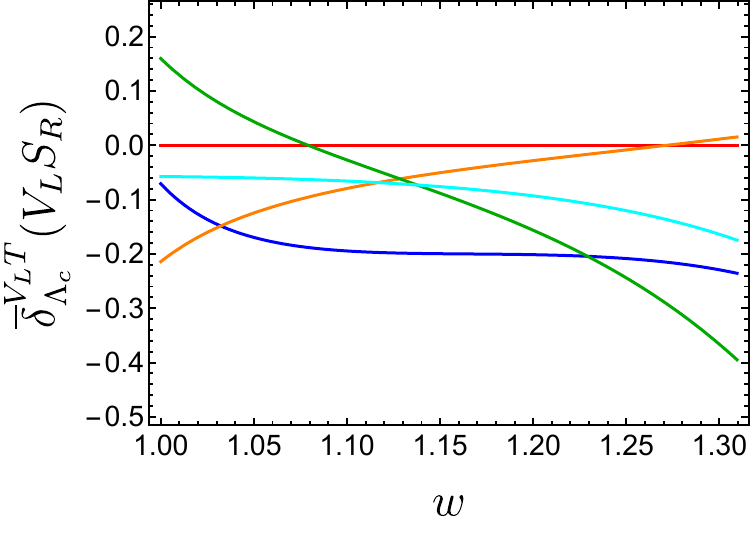}~
\includegraphics[width=0.21\linewidth]{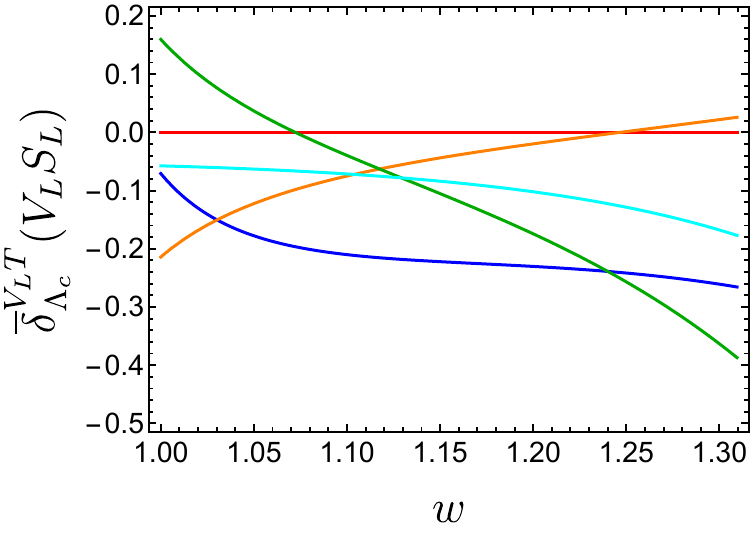}~
\includegraphics[width=0.21\linewidth]{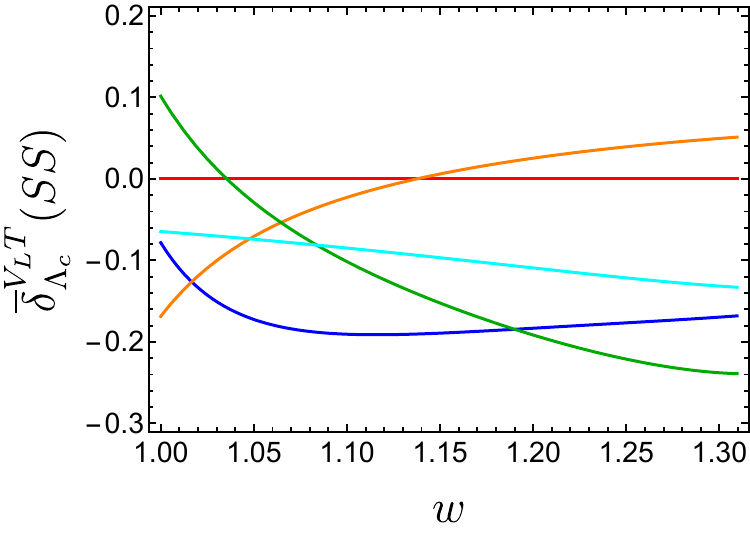}~
\includegraphics[width=0.21\linewidth]{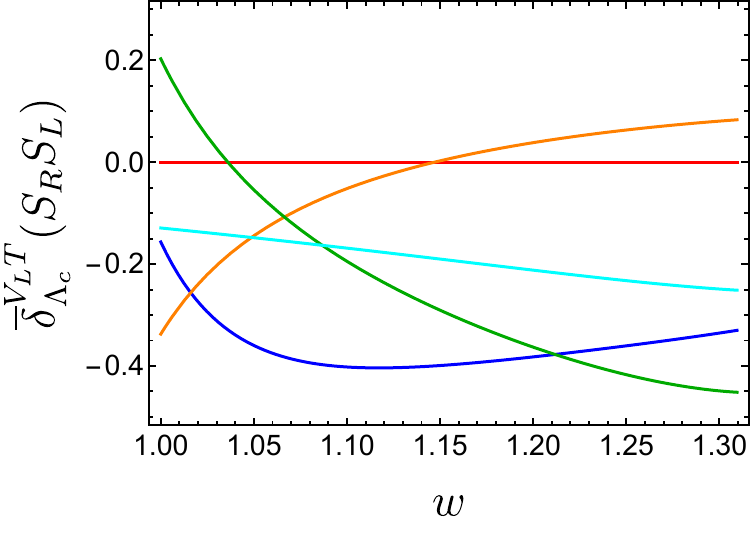}~
\includegraphics[width=0.21\linewidth]{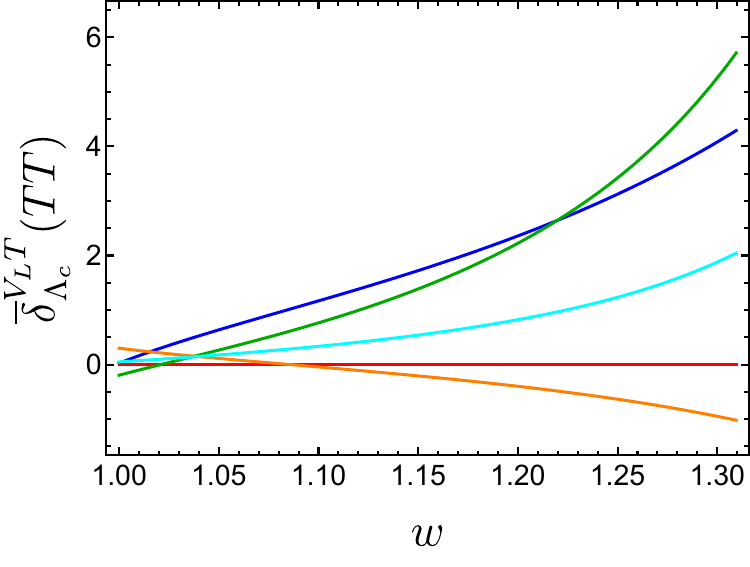}
\end{center}
\vspace{-.25cm}
\caption{$\overline{\delta}^{kl}_{\Lambda_c}(ij)$ as a function of $w$. The color-to-scenario correspondence is found in the legend of Fig.~\ref{fig:dkappaVSR} and main text. 
}
\label{fig:app_1}
\end{figure}

In Eq.~\eqref{eq:delta_mod_2}, there are several choices of which term satisfies $\overline{\delta}_{\Lambda_c}^{kl}(kl) = 0$. 
In Figs.~\ref{fig:dkappaVSR} and \ref{fig:dintVSR}, we have studied $\{kl\} = \{V_LS_R\}$. 
In this appendix, we study other cases.
In Fig.~\ref{fig:app_1}, we show $\overline{\delta}^{V_LS_L}_{\Lambda_c}(ij)$, $\overline{\delta}^{SS}_{\Lambda_c}(ij)$, $\overline{\delta}^{S_RS_L}_{\Lambda_c}(ij)$, $\overline{\delta}^{TT}_{\Lambda_c}(ij)$, and $\overline{\delta}^{V_LT}_{\Lambda_c}(ij)$.
Scenarios S1 and S2 tend to predict an opposite sign of $\overline{\delta}^{kl}_{\Lambda_c}(ij)$ to each other, and thus, the S3 prediction appears between those two results.

\begin{figure}[!t]
\begin{center}
\includegraphics[width=0.21\linewidth]{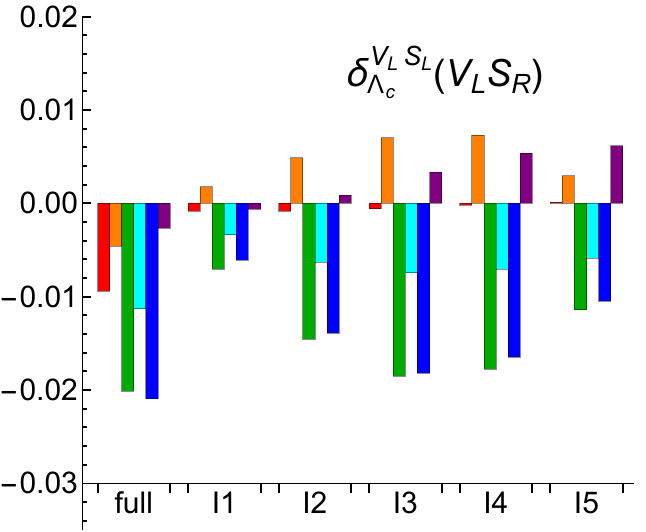}~
\includegraphics[width=0.21\linewidth]{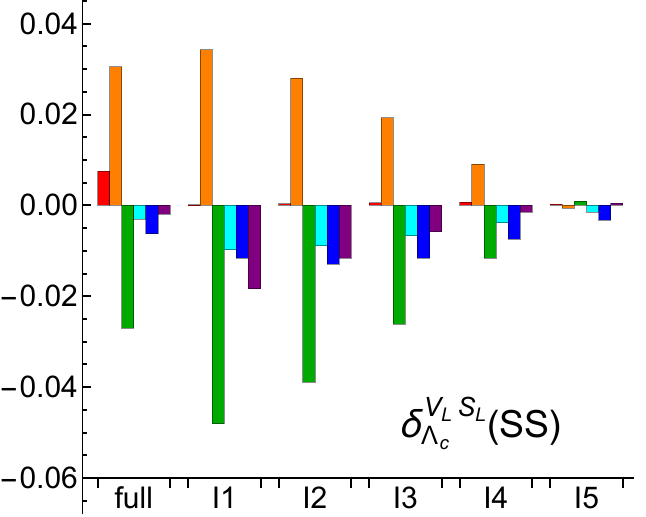}~
\includegraphics[width=0.21\linewidth]{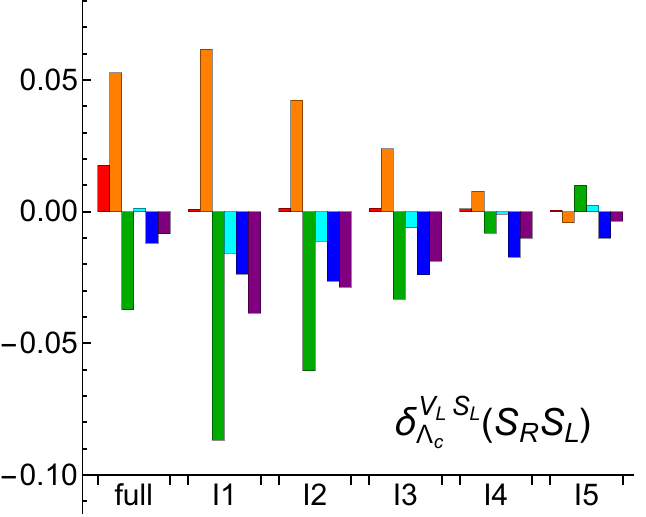}~
\includegraphics[width=0.21\linewidth]{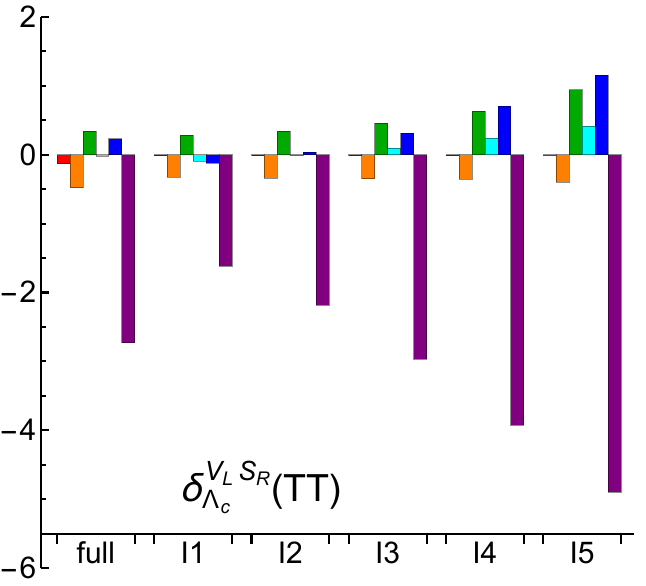}~
\includegraphics[width=0.21\linewidth]{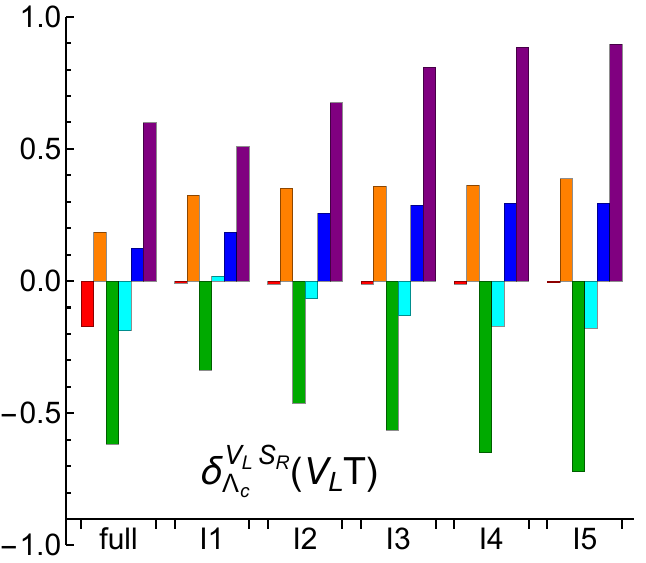}\\ \vspace{0.1cm}
\includegraphics[width=0.21\linewidth]{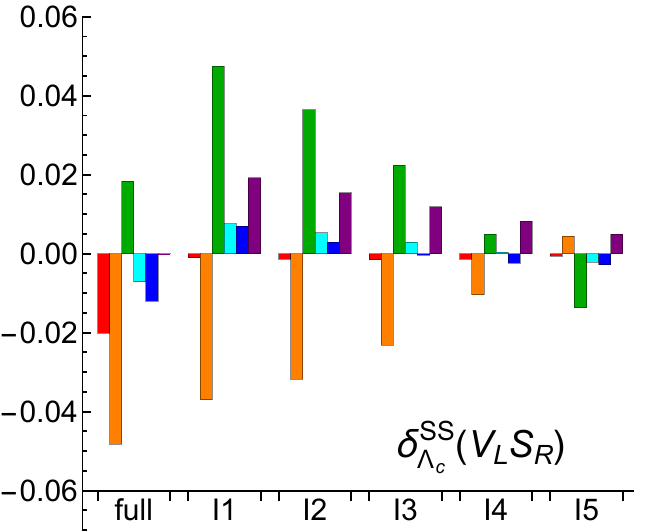}~
\includegraphics[width=0.21\linewidth]{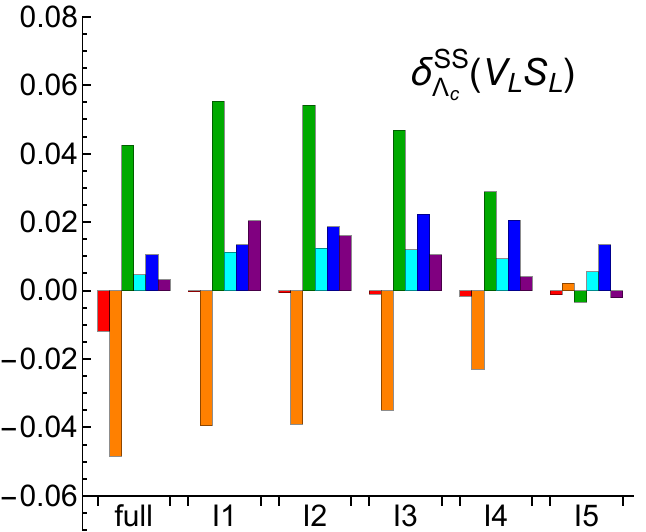}~
\includegraphics[width=0.21\linewidth]{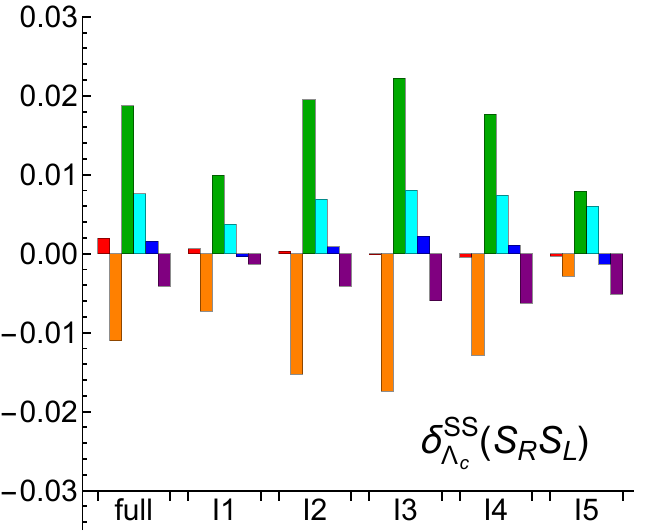}~
\includegraphics[width=0.21\linewidth]{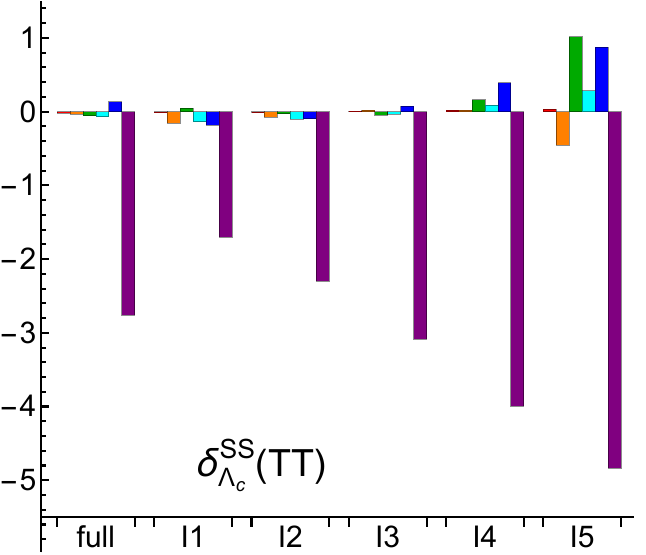}~
\includegraphics[width=0.21\linewidth]{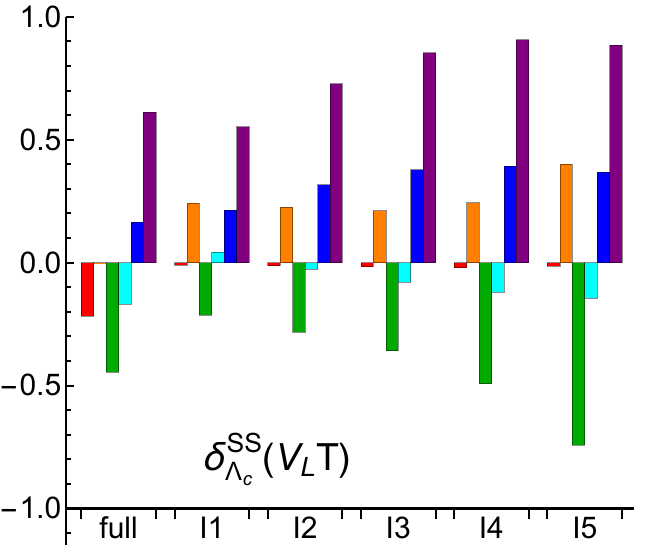}\\ \vspace{0.1cm}
\includegraphics[width=0.21\linewidth]{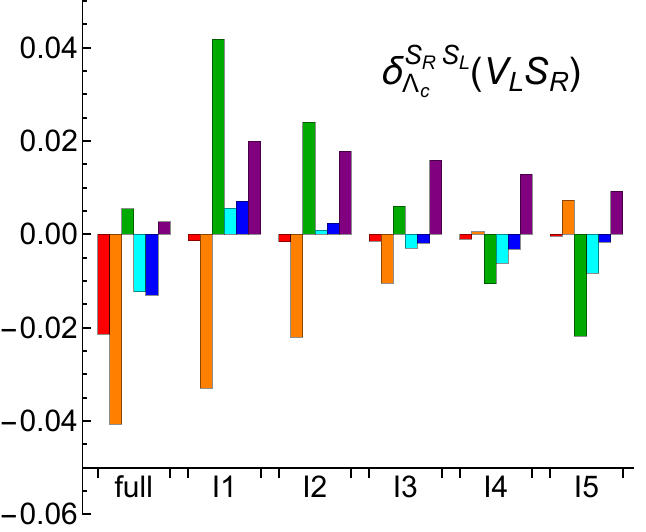}~
\includegraphics[width=0.21\linewidth]{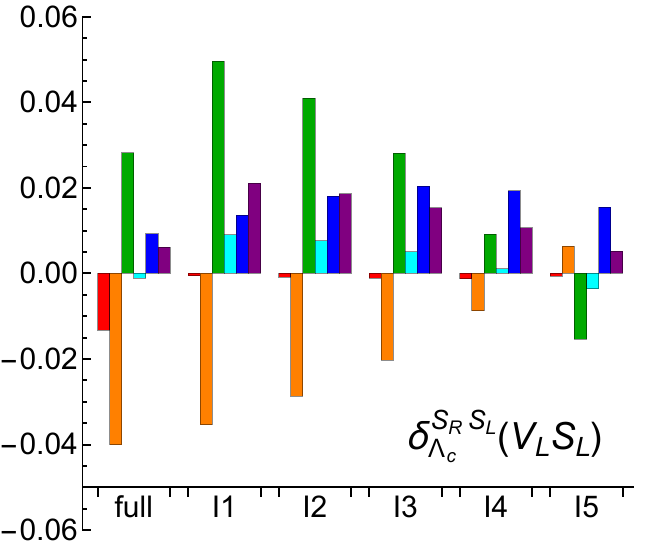}~
\includegraphics[width=0.21\linewidth]{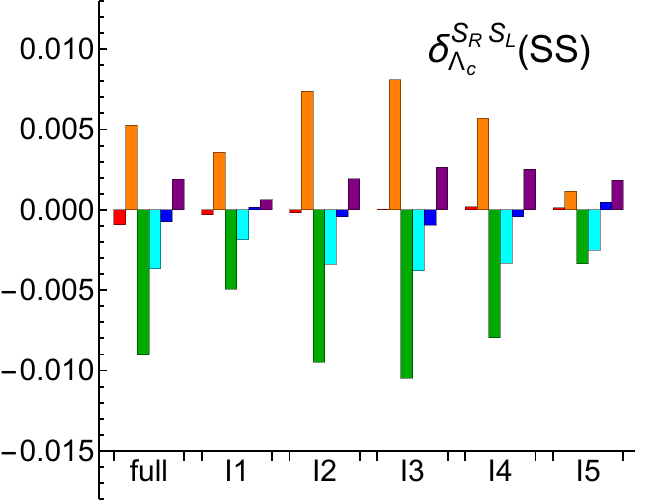}~
\includegraphics[width=0.21\linewidth]{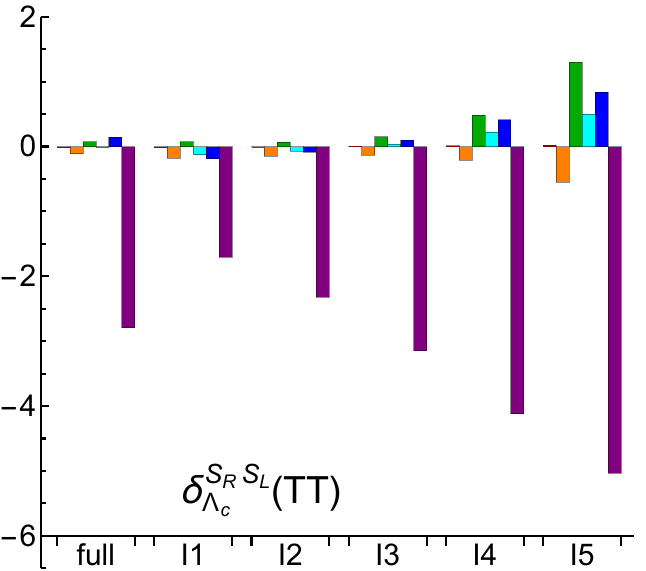}~
\includegraphics[width=0.21\linewidth]{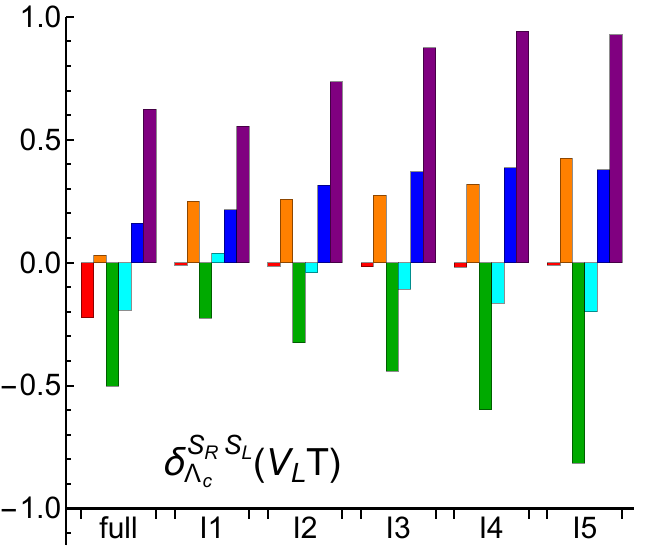}\\ \vspace{0.1cm}
\includegraphics[width=0.21\linewidth]{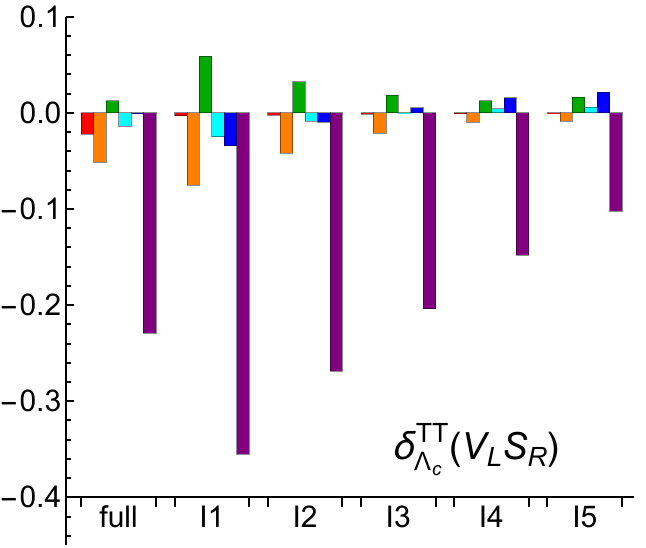}~
\includegraphics[width=0.21\linewidth]{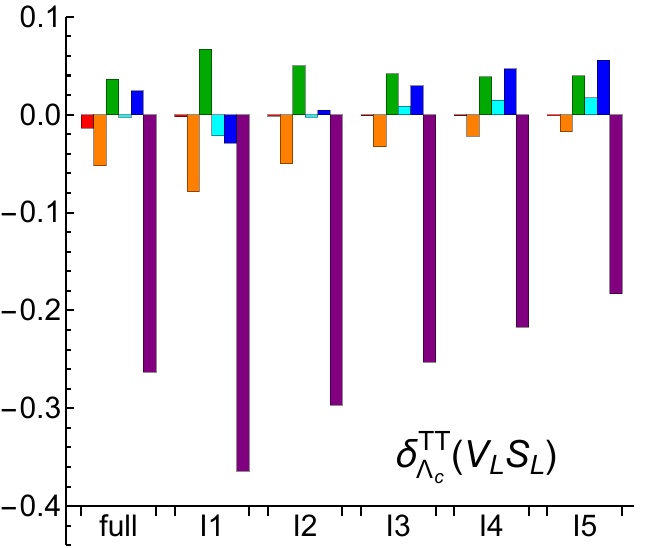}~
\includegraphics[width=0.21\linewidth]{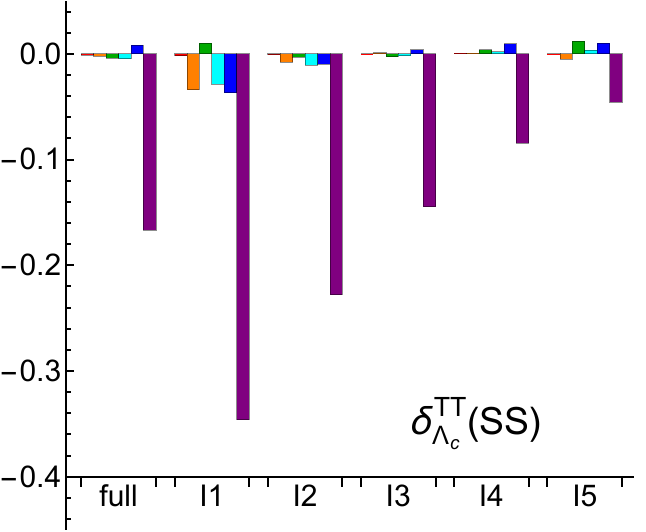}~
\includegraphics[width=0.21\linewidth]{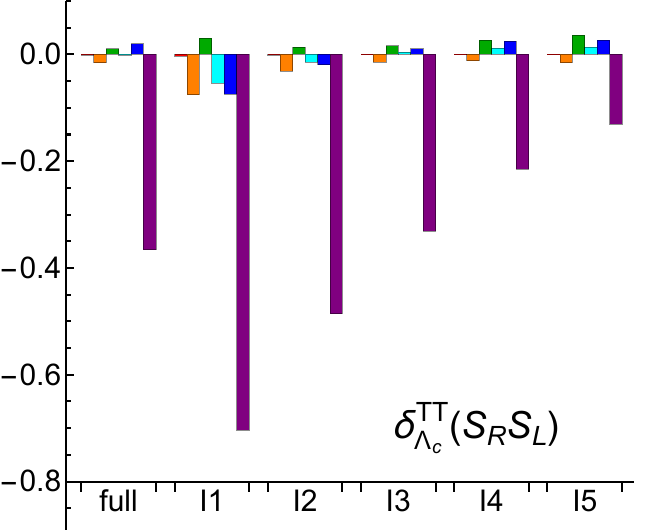}~
\includegraphics[width=0.21\linewidth]{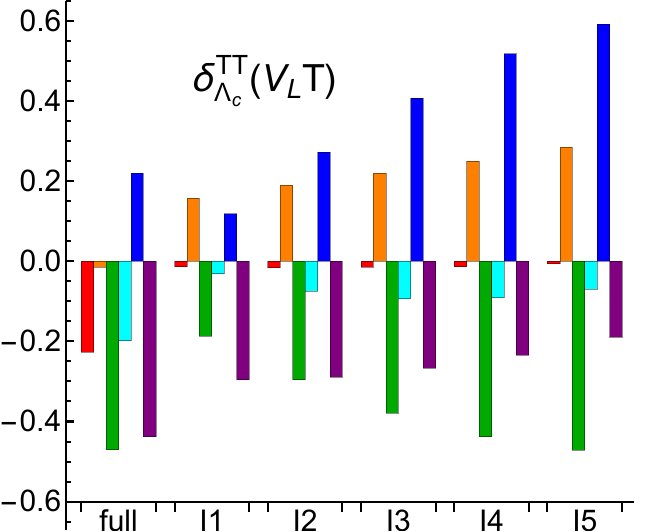}\\ \vspace{0.1cm}
\includegraphics[width=0.21\linewidth]{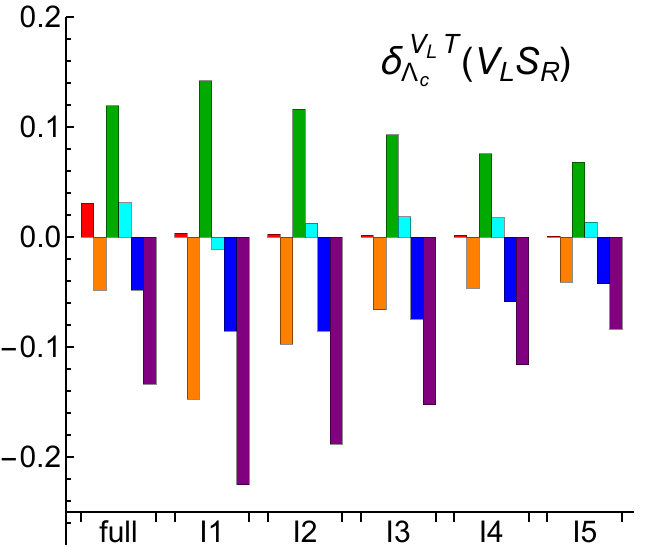}~
\includegraphics[width=0.21\linewidth]{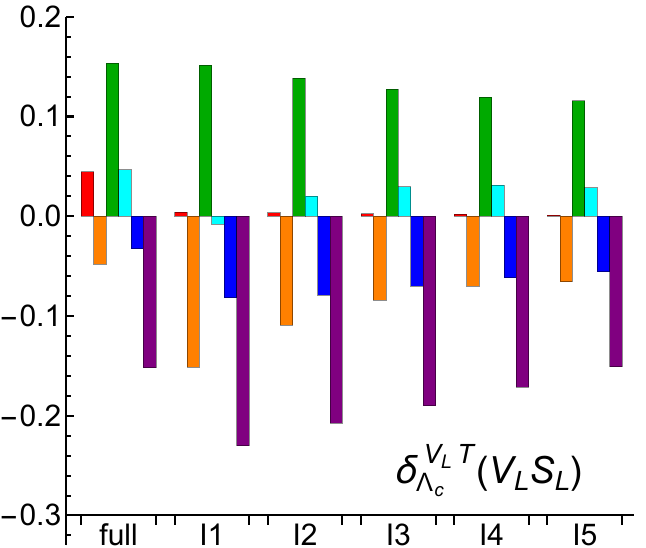}~
\includegraphics[width=0.21\linewidth]{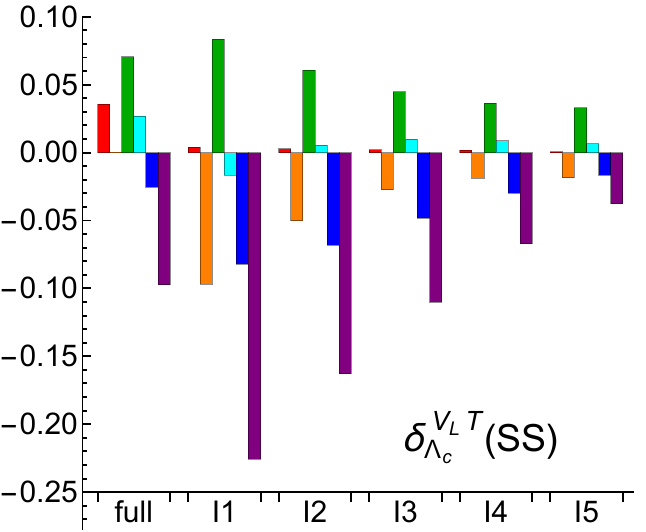}~
\includegraphics[width=0.21\linewidth]{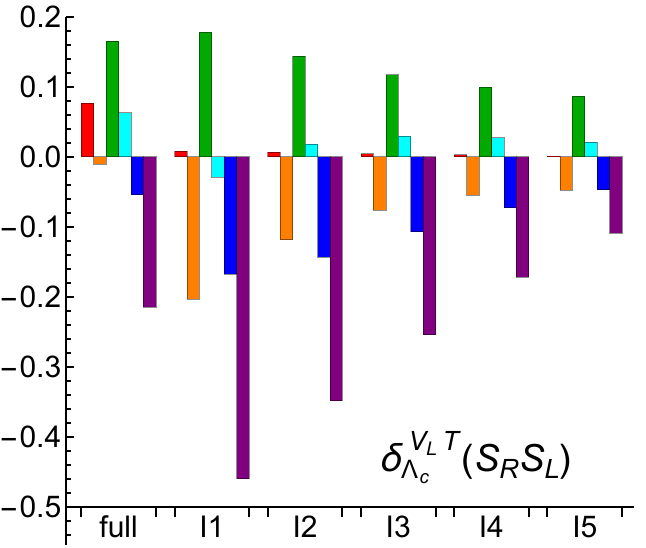}~
\includegraphics[width=0.21\linewidth]{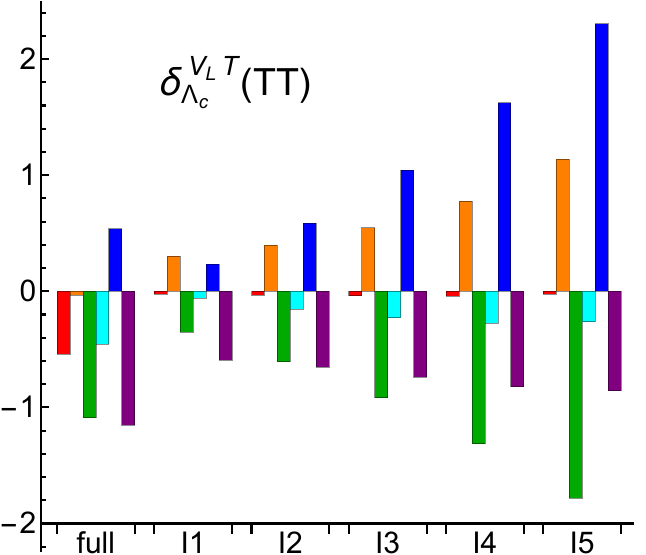}
\end{center}
\vspace{-.25cm}
\caption{$\delta^{kl}_{\Lambda_c}(ij)$ as a function of $w$.
The color-to-scenario correspondence is found in the legend of Fig.~\ref{fig:dintVSR} and main text. 
}
\label{fig:app_2}
\end{figure}

In Fig.~\ref{fig:app_2} we also show $\delta^{V_LS_L}_{\Lambda_c}(ij)$, $\delta^{SS}_{\Lambda_c}(ij)$, $\delta^{S_RS_L}_{\Lambda_c}(ij)$, $\delta^{TT}_{\Lambda_c}(ij)$, and $\delta^{V_LT}_{\Lambda_c}(ij)$. 
Among S0 -- S2, scenario S0 provides the minimum correction in most cases.
As in $\delta^{V_LS_L}_{\Lambda_c}(ij)$, those in S3 tends to be milder than the S1 and S2 cases.
In $\delta^{V_LS_L}_{\Lambda_c}(V_LS_R)$, S4 has larger values compared to S5 similarly to $\delta^{V_LS_L}_{\Lambda_c}(V_LS_L)$.
It is found that S5 tends to have the larger $\delta_{\Lambda_c}$ than the HQET results once it involves the tensor Wilson coefficient.

In scenarios S3 and S4, the deviations tend to be suppressed around $w\sim 1$.
However, this is not always the case for S5.

\bibliographystyle{utphys28mod}
\bibliography{ref}
\end{document}